\documentclass[a4paper,12pt]{article}
\pdfoutput=1
\usepackage{jheppub}
\usepackage{fixltx2e}
\usepackage{float}
\usepackage{graphicx,amssymb,bm,latexsym,amsmath}
\usepackage{subfigure,float,psfrag,rotating}
\usepackage{epstopdf}
\usepackage[section]{placeins}
\usepackage{capt-of}
\usepackage{caption}
\newcommand{\sn}{{\rm sn}}
\newcommand{\cn}{{\rm cn}}
\newcommand{\dn}{{\rm dn}}

\usepackage[toc,page]{appendix}

\title{\boldmath Perturbations of giant magnons and single spikes in $\mathbb R \times S^2$}
\author{Soumya Bhattacharya${}^{a}$,}
\author{Sayan Kar${}^{b}$,} 
\author{Kamal L. Panigrahi${}^{b}$} 
\affiliation{${}^{a}$ Department of Theoretical Sciences\\
S. N. Bose National Centre 
for Basic Sciences, JD Block, Sector III, Salt Lake City, Kolkata -700 098, India}
\affiliation{${}^{b}$Department of Physics,\\Indian Institute of Technology Kharagpur,\\
Kharagpur-721 302, India}
\emailAdd{soumya557@bose.res.in, soumya557@gmail.com} 
\emailAdd{sayan@phy.iitkgp.ac.in}
\emailAdd{panigrahi@phy.iitkgp.ac.in}

\abstract{Perturbations of giant magnons and single spikes in a $2+1$ dimensional $\mathbb R \times S^2$ background spacetime are analysed. Using the form of the
giant magnon solution in the Jevicki-Jin gauge,
the well-known Jacobi equation for small normal deformations of an embedded time-like surface are written down. Surprisingly, this equation reduces to a simple wave equation in a Minkowski background. The finiteness of perturbations and the ensuing stability of such giant magnons under small deformations are then discussed. It turns out that only the zero mode has finite deformations and is stable.  Thereafter, we move on to explore the single spike solution in the Jevicki-Jin gauge. We  obtain
and solve the perturbation equation numerically and  address stability issues. }

\keywords{Giant magnons, Single spikes, perturbations}

\begin{document}

\maketitle
  \flushbottom
\section{Introduction}

 Rigidly rotating strings with cusps or spikes have been investigated earlier quite extensively in the context of cosmic strings \cite{BurdenTassie1: 1982, BurdenTassie2: 1982,BurdenTassie3: 1984,Burden: 1985,Embacher: 1992,rigid:1,rigid:2,rigid:3,Burden: 2008, ogawa:2008}. In its modern avatar, such rigidly rotating strings have been
renamed as {\em spiky strings}. Their importance and relevance today 
is largely in the context of  the AdS/CFT correspondence
\cite{Maldacena:1997re}. More than a decade ago, the spiky string solutions
appeared through the seminal work of Kruczenski
\cite{Kruczenski:2004wg}, as a potential gravity dual to higher
twist operators in string theory. 
The semiclassical aspect of these solutions arise through
evaluating their energies, angular momenta and finding relations between them \cite{Gubser:2002tv}.  
\noindent 
In providing meaningful input towards the realization of the gauge-gravity duality, the
emergence of integrability on both sides have been quite useful and important. 
Classical and quantum integrability of the ${\cal N}$ = 4 Supersymmetric Yang-Mills (SYM) theory in the
planar limit is mainly useful for the remarkable advances in understanding the theory \cite{Minahan:2002ve, dolan, nappi}. The
nonlocal conserved charges found on the string side \cite{Mandal:2002fs, Bena:2003wd} appear to have a counterpart in planar
gauge theory at weak coupling within the spin-chain formulation for the dilatation operator.
 In this connection, Hoffman and
Maldacena \cite{HM} considered a special limit $(J\rightarrow \infty$,
$\lambda$ = fixed, $p$ = fixed, $E-J$ = fixed), with $J$ being the
angular momentum, $E$ the total energy, $p$ the momentum carried
by the string in the spin chain and $\lambda$  the `t Hooft
coupling constant. They considered operators of the form $O_p \sim \sum_l e^{ilp} (...ZZZWZZZ...)$, where the field $W$ is inserted at position $l$ along the spin-chain. In this limit, 
the problem of determining the
spectrum on both (i.e. string and gauge theory) sides becomes
simple. Such impurities are called ``magnons'' that propagate along the spin-chain with a conserved momentum $p$. In the large $\lambda$ limit the dispersion relation can be written in the form
\begin{equation*}
E - J = \frac{\sqrt{\lambda}}{\pi}|\sin{\frac{p}{2}}| \, \, .
\end{equation*}
To describe these elementary excitations in the string theory side  the authors in \cite{HM} considered the string sigma model in AdS${}_5$ $\times$
S ${}^5$ and have taken the large $\lambda$ limit. In this limit,
the classical approximation is valid. Computing energy in this limit they found
\begin{equation*}
E - J = \frac{\sqrt{\lambda}}{\pi}\sin{\frac{\Delta \phi}{2}} \, \, ,
\end{equation*}
where $\Delta \phi$ represents a geometric angle in the string theory side. Now identifying $\Delta \phi = p$, one can have the perfect match between the two dispersion relations.
\noindent Giant magnons can be visualised as a special limit of such spiky
strings with shorter wavelengths. In \cite{ishizeki07} another 
 limit called the single spike limit of rigidly rotating string the solution can be visualized as a string wound infinitely along the equator of $S^2$ with a single spike pointing towards the north pole. A large class of such spiky strings and giant magnons in various backgrounds have been studied, for example, in
\cite{Kruczenski:2004wg},\cite{Frolov:2003qc,dorey1,chen1,chen2,russo,hirano,hofman07,swanson,ishizeki07,arutyunov,dorey2,ishizeki08,Biswas:2012wu,Banerjee:2014gga,Banerjee:2015nha}. \noindent Given the above-stated importance of the spiky strings and giant magnons, it is  
worthwhile to look at the geometric
properties of such string configurations from the world-sheet view
point. Here following our earlier works  \cite{bkp:2017}, \cite{bkp:2018} we study normal deformations (linearized) about the
classical giant magnon solution in $\mathbb R \times S^2$ using the well-known Jacobi equations
\cite{garriga: 1993,guven: 1993,frolov: 1994,capovilla: 1995} which govern normal deformations of an
embedded surface. Earlier work in \cite{Frolov:2002av} dealt with
computation of quantum corrections to the energy spectrum, 
by expanding the supersymmetric action to quadratic order in
fluctuations about the classical solution. The linearized perturbations of semiclassical
strings are extremely instructive in matching the duality beyond the leading order 
classical solutions.  The main motivations behind the perturbative solutions are multi-fold. On one hand, it helps in studying the stability properties
of the string dynamics of the closed strings, and finding the quantum string corrections to 
the Wilson loop expectation value for the open string solutions and on the other hand, it helps us in determining the physical properties of topological defects. 
\noindent  Apart from \cite{bkp:2017, bkp:2018}, related recent works on such perturbations can be found
in \cite{forini:2015, rojas, ida, barik}.
The rest of this article is organised as follows. In Section 2, we
briefly summarize the two different embeddings
of the worldsheet and the rigidly rotating string solution in $\mathbb R \times S^2$. 
Section 3 is devoted to a study of the normal perturbations of the giant magnons in $\mathbb R \times S^2$. Section 4 is devoted to the single spike case.
Our final concluding remarks appear in Section 5. Finally in the appendix we have briefly discussed the algorithm of finite difference method which we have used extensively in section 4.

\section{Rigidly rotating strings in $\mathbb R \times S^2$:
Kruczenski and Jevicki-Jin embeddings}

\noindent The bosonic string  worldsheet embedded in a 
$N$ dimensional curved spacetime with background metric
functions $g_{ij}(x)$, is described by the well-known Nambu-Goto action given as,
\begin{equation}
S =-T \int d\tau \,\, d\sigma \sqrt{-\gamma} = - T \int  d\tau \,\, 
d\sigma \sqrt{ (\dot{x}\cdot x')^2-\dot{x}^2 {x'}^2} \ .
\label{ac1}
\end{equation}
Here $\gamma$ denotes the determinant of the induced metric $\gamma_{ab}=g_{ij}
\partial_a x^{i} \partial_b x^{j}$ ($a,b=\sigma,\tau$). 
The $x^{i}(\tau,\sigma)$ are functions which describe the
profile of the string worldsheet, as embedded in the target spacetime. Here
$T$ denotes the string tension. We have used the notation: $\dot x = \partial_{\tau} x$,\,
$ x' = \partial_{\sigma} 
x$, \, $(\dot x \cdot x') = g_{ij}{\dot x}^{i} \, {x'}^{j}$,\, 
${\dot x}^2=g_{ij}{\dot x}^{i} \,{\dot x}^{j}$ and
${x'}^2=g_{ij}{ x'}^{i} \,{x'}^{j}$.

\noindent Let us now turn to rigidly rotating strings in $\mathbb R \times S^2$ background. The three dimensional background line element is given as,
\begin{equation}
ds^2 = -dt^2 + d\theta^2 + \sin^2 \theta d\phi^2 \, \, .
\end{equation}
\noindent This $2+1$ dimensional spacetime is non-flat but has constant Ricci curvature. The value of $R$ is given as 
\begin{equation}
R = 2 \, .
\end{equation}
\noindent However this is not Einstein space for which one has 
\begin{equation}
R_{ij} = \Lambda g_{ij} \, \, .
\end{equation}
We may refer to the background space as one with constant positive curvature.

\noindent The embedding used in \cite{ishizeki07} is as follows,
\begin{equation}
 t= \kappa \tau , \ \ \theta = \theta(\sigma), \ \ \phi = \omega \tau + \sigma \, \, .
\label{ansz1}
\end{equation}
Here $\omega$ and $\kappa$ are two parameters.

\noindent From the equations of motion one can arrive at the following 
first integral for $\theta$.
\begin{equation}
\theta' = \frac{\kappa \sin \theta}{C}~\sqrt{\frac{\omega^2 \sin^2 \theta - C^2}{\kappa^2 - \omega^2 \sin^2 \theta}} \, \, ,
\label{th}
\end{equation}
where $C$ is the integration constant.

\noindent As discussed in \cite{ishizeki07} equation (\ref{th}) can have two distinct limits: 1) $|\omega| \rightarrow \kappa$ and 
~2) $|\omega| \rightarrow C$. Both these limits simplify the mathematical expressions. The first of these limits will give rise to the giant magnon solution which is identified by the following
(magnon) dispersion relation,
\begin{equation}
E - J = \frac{\sqrt{\lambda}}{\pi}~ |\sin {\frac{\Delta \phi}{2}}| \, \, .
\end{equation}
In the dual field theory the $\Delta \phi$ is mapped onto momentum--hence
the name `magnon'.  

\noindent The second limit will give rise to what is called the single spike solution and described by a string wrapped around the equator, infinite number of times with a single spike 
pointing towards the pole of the $2$-sphere. Here we will focus on the giant magnon solution. 

\noindent  Here  we work in the conformal gauge and use the embedding 
due to Jevicki and Jin \citep{jevicki08}, given as 
\begin{equation}
t= \kappa \tilde \tau + f(\tilde \sigma) , \hspace{0.1in} \theta=
\theta(\tilde \sigma) , \hspace{0.1in} \phi= \omega \tilde \tau +
g(\tilde \sigma) \, \, .
\end{equation}
\noindent where $\tilde \tau$ and $\tilde \sigma$ are the worldsheet coordinates. We retain $\kappa$, $\omega$ as in the embedding mentioned before in 
equation (\ref{ansz1}).

\noindent In conformal gauge, 
we have the following Virasoro constraint equations
\begin{equation}
g_{ij}({\dot x}^{i}{\dot x}^{j}+{x'}^{i}{x'}^{j}) = 0, \hspace{0.3in} g_{ij}{\dot x}^{i}{x'}^{j} = 0 \ .
\label{ee2}
\end{equation}

\noindent In such a gauge, the string equations of motion 
obtained by varying the action with respect to $x^i$, take the form:
\begin{equation}
{\ddot x}^{i} - {x^{i ''}} + \Gamma^{i}_{jk} \left
( {\dot x}^{j} {\dot x}^{k} - {x}^{j \prime} {x}^{k\prime}
\right )  = 0 \ .
\label{ee1}
\end{equation}

\noindent Using the above ansatz and after some simplifications, we have the following equation for $\theta(\tilde \sigma)$,
\begin{equation}
\theta' =  \pm \sqrt{(\kappa^2 + a^2) - (\omega^2 \sin^2 \theta + \frac{C^2}{\sin^2 \theta})} \, \, ,
\label{th1}
\end{equation}
which also satisfies one of the Virasoro conditions. Note that the prime on the  L. H. S. above is w.r.t. $\tilde \sigma$.

\noindent From the other Virasoro condition (\ref{ee2}) we have 
\begin{equation}
\kappa ~a = \omega ~C \, \, .
\label{vir1}
\end{equation}
Using this equation for replacing the parameter $a$, we 
rewrite the equation for $\theta$ as 
\begin{equation}
\theta' =  \pm \frac{1}{\kappa \sin \theta} 
\sqrt{(\kappa^2 - \omega^2 \sin^2\theta)(\kappa^2 
\sin^2 \theta - C^2)} \, \, 
\end{equation}

\subsection{Giant magnons in $\mathbb R \times S^2$}

\noindent In the giant magnon limit $|\omega| \rightarrow \kappa$, from equation (\ref{vir1}) we have 
\begin{equation}
a= C \, \, .
\end{equation} 

\noindent Equation (\ref{th1}) will take the following form
\begin{equation}
\theta' = \pm \frac{\cos \theta}{\sin \theta}\sqrt{\omega^2 \sin^2 \theta - C^2} \, \, .
\label{th2}
\end{equation}

\noindent The equations of motion and the Virasoro constraints give us the following conditions,
\begin{eqnarray}
f'(\tilde \sigma) = a \hspace{0.2in}
; \hspace{0.2in} g'(\tilde \sigma) = \frac{C}{\sin^2 \theta} \, \, .
\end{eqnarray}

\noindent Integrating equation (\ref{th2}) we have the expression for $\theta$
\begin{equation}
\cos \theta = \frac{\alpha}{\cosh \alpha \omega \tilde \sigma}
\label{ths1}
\end{equation}

where $\alpha^2 = 1 - \frac{c^2}{\omega^2}$. Similarly $f(\tilde \sigma)$ and $g(\tilde \sigma)$ will be of the following form
\begin{eqnarray}
f(\tilde \sigma) &=& C \tilde \sigma + b \, \, , \\
g(\tilde \sigma) &=& C \tilde \sigma \mp \tan^{-1} 
\left(\frac{\sqrt{1-\alpha^2}}{\alpha} \coth \, \omega \alpha 
\tilde \sigma \right)
\end{eqnarray}
Without loss of generality we can set $b = 0$.

\noindent The relation between the two embeddings can be understood as follows. For embedding (\ref{ansz1}) the induced metric is not diagonal while for the conformal gauge it is so. The 
$\tau, ~\sigma$ in the non-diagonal case are related to those in the diagonal case as follows:
\begin{equation}
\tau = \tilde \tau + \frac{f(\tilde \sigma)}{\omega}, \, \, \, \, \hspace{0.2in} \sigma = g(\tilde \sigma) - f(\tilde \sigma) \, \, .
\end{equation}

\noindent We can transform the derivative in the  L. H. S of the equation for $\theta'$ 
quoted above into a derivative w.r.t. $\sigma$ by using the above relation.
It is easy to check that the equation in the Jevicki-Jin gauge goes over
to that in the Kruczenski gauge. Further, using the relation between
$\sigma$ and $\tilde \sigma$ one can convert the solution for $\theta(\tilde\sigma)$ to the solution $\theta (\sigma)$ in the Kruczenski gauge. It is important to realise that while the $\sigma$ runs from $\theta_0$ to $\frac{\pi}{2}$
(or $-\theta_0$ to $-\frac{\pi}{2}$, $\sin \theta_0 = \frac{C}{\omega}$),
the $\tilde \sigma$ runs from $0$ to $\pm \infty$. The values of the
spacetime coordinate $\theta$ in both gauges run from $\theta_0$ to
$\frac{\pi}{2}$ (or their negative valued counterparts). 

\noindent  We now turn towards
examining the stability of the above string configurations by studying their normal
deformations.

\section{Perturbations and stability of giant magnons in $\mathbb R \times S^2$}
Before investigating the perturbation equations for
our specific solutions, let us briefly recall the well-known Jacobi
equations which deal with
perturbations of extremal worldsheets.

\subsection{Jacobi equations for extremal surfaces}
Given that $x^i(\tau,\sigma)$ are the embedding functions and
$g_{ij}$ the background metric, the tangent vectors to the worldsheet are
\begin{equation}
e^i_{\tau} = \partial_\tau x^i , \hspace{0.2in}
e^i_{\sigma} = \partial_\sigma x^i ,
\end{equation}
Thus, the induced line element turns out to be,
\begin{equation}
\gamma_{ab} = g_{ij} e^i_a e^j_b \ ,
\end{equation}
where the $a,b...$ denote worldsheet indices (here $\tau$,
$\sigma$). The worldsheet normals $n^i_{(\alpha)}$ satisfy
the relations
\begin{equation}
g_{ij} n^i_{(\alpha)}n^{j}_{(\beta)} = \delta_{\alpha\beta}
 ,\hspace{0.2in} g_{ij}n^i_{(\alpha)} e^j_{a}=0 \ ,
\end{equation}
where $\alpha=1..,N-2$ and $N$ is the dimension of the
background spacetime. The last condition holds for all $\alpha$
and $a$. Extrinsic curvature tensor components $K_{ab}^{(\alpha)}$ along
each normal $n^i_{(\alpha)}$ of the embedded worldsheet are
\begin{equation}
K_{ab}^{(\alpha)} = -g_{ij}( e^k_{a}\nabla_k e^i_b) n^{j(\alpha)}
\ .
\end{equation}
The equations of motion lead to the
condition, $K^{(\alpha)} = \gamma^{ab} K_{ab}^{(\alpha)} =0$
for an extremal worldsheet.
Thus extremal surfaces are those for which the trace of the extrinsic curvature tensor
along each normal is zero. Normal
deformations are denoted as
$\phi^{(\alpha)}$ along each normal. Therefore, the deformations constitute a
set of scalar fields. More explicitly, the deformation of each coordinate is
\begin{equation}
\delta x^i = \phi^{(\alpha)} n^i_{(\alpha)} \ ,
\end{equation}
which is the perturbation of the worldsheet (i.e. $x^i
\rightarrow x^i + \delta x^i$). For a worldsheet
with $\gamma^{ab}K_{ab}^{(\alpha)}=0$ satisfying the equations of
motion and the Virasoro constraints, the scalars
$\phi^i_{(\alpha)}$ satisfy the Jacobi
equations given as.
\begin{equation}
\frac{1}{\Omega^2} \left ( - \frac{\partial^2}{\partial \tau^2} +
\frac{\partial^2}{\partial \sigma^2} \right ) \phi^{(\alpha)} +
\left (M^2\right )^{(\alpha)}_{(\beta)} \phi^{(\beta)} = 0 \ ,
\end{equation}
where
\begin{equation}
\left (M^2\right )^{(\alpha)}_{(\beta)} = K_{ab}^{(\alpha)}K^{ab}_
{(\beta)} +R_{ijkl} e^j_a e^{l\,\,a} n^{i\,\,(\alpha)}
n^k_{(\beta)} \ ,
\end{equation}
and $\Omega^2(\tau,\sigma)$ is the conformal factor of the
conformally flat form of the worldsheet line element. The Jacobi equations are
obtained by constructing the second variation of the worldsheet action. 
Thus, solving these equations
for the perturbation scalars one can analyse the stability properties
of the extremal worldsheet. In other words, a stable worldsheet will
correspond to an oscillatory character for the $\delta x^i$ defined above. 

\noindent  Note further that the Jacobi
equations are like a family of coupled, variable `mass'  wave
equations for the scalars $\phi^{(\alpha)}$. They are 
a generalisation of the familiar geodesic deviation equation for 
geodesic curves in a Riemannian geometry. Usually, these equations 
are quite complicated and not easily solvable, even for the
simplest cases. It turns out that for the string configurations under consideration here,
we do find analytical solutions.

\noindent In a more general context, the
worldsheet covariant derivative
does have a term arising 
from the extrinsic twist potential (normal fundamental form)
which is given as:
$\omega_a^{\,\,\alpha\beta}=
g_{ij}\, \left ( e^k_a\nabla_k \, n^{i\,\alpha}\right )\,n^{j\,\beta}$.
For codimension one surfaces, i.e. hypersurfaces (as is the case here), the $\omega_a^{\alpha\beta}$
are all identically zero.

\subsection{The case of giant magnons in $\mathbb R \times S^2$}

\noindent We will only look at the giant magnon configurations
in the $\mathbb R \times S^2$ and compute the perturbations. To proceed, we 
first write down the normal and the
extrinsic curvature for the world sheet configurations, in the
Jevicki-Jin gauge stated earlier. Below, we will use $\tilde \sigma$, $\tilde \tau$ for the
Jevicki-Jin gauge. 

\noindent The induced metric is given as:
\begin{equation}
ds^2 = \omega^2 \cos^2 \theta \left( -d{\tilde\tau}^2 + d{\tilde \sigma}^2 \right) =  \frac {\omega^2\alpha^2}{\cosh^2 {\tilde \sigma}}~\left(-d{\tilde \tau}^2 + d
{\tilde \sigma}^2 \right) \, \, .
\end{equation}

\noindent The tangent vectors to the worldsheet are given as:
\begin{equation}
e^{i}_{\tilde\tau} = \left (\omega, ~0, ~\omega\right ) \hspace{0.2in}; \hspace{0.2in}
e^{i}_{\tilde \sigma} = \left ( f', ~\theta', ~g'\right ) \, \, .
\end{equation}

\noindent The normal to the worldsheet is given as:
\begin{equation}
n^{i} = \left( \frac{\sqrt{\omega^2 \sin^2 \theta - C^2}}{\omega \cos \theta}, ~-\frac{C}{\omega \sin \theta}, ~\frac{\sqrt{\omega^2 \sin^2 \theta - C^2}}{\omega \sin^2 \theta \cos \theta} \right)
\label{nrm1}
\end{equation}

\noindent The extrinsic curvature tensor which is defined as $K_{ab}^{(\alpha)} = -g_{ij}( e^k_{a}\nabla_k e^i_b) n^{j(\alpha)}$, turns out to be
\begin{equation}
K_{ab} = \begin{pmatrix} -C \omega \cos \theta & -\omega^2 \cos \theta \cr -\omega^2 \cos \theta & -C \omega \cos \theta \end{pmatrix} .
\end{equation}
\noindent Note that the trace of $K_{ab}$ vanishes (i.e $\gamma^{ab} K_{ab} = 0$), which is the classical equation of motion.
\noindent Hence the quantity $K_{ab}K^{ab}$ is given as:
\begin{equation}
K_{ab}K^{ab} = \frac{2(C^2 - \omega^2)}{\omega^2 \cos^2 \theta} \, \, .
\end{equation}

\noindent Similarly one can write down the contribution from the
Riemann tensor term in the perturbation equation as follows:
\begin{equation}
R_{ijkl} ~e^j_a ~e^{l\,\,a} ~n^{i\,\,(\alpha)}~n^k_{(\beta)} = \frac{(\omega^2 - C^2)}{\omega^2 \cos^2 \theta}\delta^{\alpha}_{\beta} \, \, .
\end{equation}

\noindent Using all of the above-stated quantities which appear
in the perturbation equation and after some lengthy algebra, 
one arrives at the
following simple equation for the perturbation scalar $\phi$
\begin{equation}
\left( -\partial_{\tilde\tau}^2 + \partial_{\tilde\sigma}^2 \right) \phi + \left( C^2 - \omega^2 \right) \phi = 0 \, \, .
\label{pert1} 
\end{equation}

\noindent Let us now use the ansatz 
\begin{equation}
\phi(\tilde \tau, \tilde \sigma) = \epsilon e^{i \beta \tilde\tau}~P(\tilde\sigma) \, \, ,
\label{ansz}
\end{equation}
\noindent where $\beta$ is the eigenvalue and $\epsilon$, a constant which we may relate to the
amplitude of the perturbation. It must be emphasized that $\epsilon$
has to be small in value (i.e. $\epsilon<<1$) in order to ensure that the
deformation is genuinely a small perturbation.

\noindent Equation (\ref{pert1}) will take the following form after using the above ansatz
\begin{equation}
\frac{d^2P}{d \tilde\sigma^2} + \left( \beta^2 + C^2 - \omega^2 \right)P = 0 \, \, .
\label{pert2}
\end{equation}

\noindent Depending on the sign of $\beta^2 + C^2 - \omega^2$,  
equation (\ref{pert2}) can have oscillatory or exponential general solutions 
given as
\begin{eqnarray}
P_1(\tilde \sigma) &=& A \cos M \tilde \sigma + B \sin M \tilde \sigma \\ 
P_2(\tilde \sigma) &=& A e^{-M \tilde \sigma} + B e^{M \tilde \sigma} 
\label{p} 
\end{eqnarray}
where $M^2 = \beta^2 + C^2 - \omega^2 = \beta^2 - \alpha^2 \omega^2$. Now $M^2 > 0$ is for the oscillatory
solution and $M^2<0$ is for the exponential solution.

\noindent Thus, our complete solution will be of the following form
\begin{equation}
\phi(\tilde \tau,\tilde \sigma) = \epsilon~e^{i \beta\tilde \tau} P_{1,2} \, \, 
\end{equation}
where, one will have to choose either $P_1$ or $P_2$.

\noindent Using the solution for $\theta$ (equation (\ref{ths1})) we can write the equation for the normal $n^{i}$ (equation (\ref{nrm1})) as
\begin{equation}
n^{i} = \left( \sinh \bar \sigma, ~ -\frac{C}{\omega} \frac{\cosh \bar \sigma}{\sqrt{\cosh^2 \bar \sigma - \alpha^2}}, ~ \frac{\sinh \bar \sigma \cosh^2 \bar \sigma}{\cosh^2 \bar \sigma - \alpha^2} \right) \, \, .
\end{equation}
\noindent Here $\bar \sigma = \alpha~\omega~ \tilde \sigma$.

\noindent Notice that the $n^0$ and $n^2$ have an overall $\sinh\bar \sigma$
which diverges as $\bar\sigma$ tends to infinity. Now if we consider the solution $P_1(\tilde \sigma)$ one can clearly see that $P_1(\tilde \sigma) \sinh \bar \sigma$ will grow exponentially with $\tilde \sigma$ and thus we end up with an unstable solution. Hence one needs to look for the other solution $P_2(\tilde \sigma)$. One can see from equation (\ref{p}) that $P_2(\tilde \sigma)$ has two terms, one is exponentially decaying with $\tilde \sigma$ while the other one is exponentially growing with $\tilde \sigma$. Since we are working here with an open string worldsheet, $\tilde \sigma$ runs from $0$ to $\infty$. We can impose a boundary condition on $\phi(\tilde \tau, \tilde \sigma)$ by considering that as $\tilde \sigma$ goes to $\infty$ the perturbation will vanish i.e: 
 $\phi(\tilde \tau, \tilde \sigma) \rightarrow 0$ as $\tilde \sigma \rightarrow \infty$ which should follow from  $P_2(\tilde \sigma) \rightarrow 0$ (at $\tilde \sigma \rightarrow \infty$). To have $P_2(\tilde \sigma) \rightarrow 0$ at $\tilde \sigma \rightarrow \infty$ we require $B = 0$. Thus, we can safely 
choose the exponentially damped solution, i.e.  $P_2 \propto 
e^{- M \tilde \sigma}$. Multiplying this $P_2$ by $\sinh \bar \sigma$
would result in

\begin{equation}
P_2 \sinh \bar \sigma = \frac{1}{2}e^{-M\tilde \sigma} \left (e^{\alpha \omega \tilde \sigma} - e^{-\alpha \omega \tilde \sigma} \right  ) 
= \frac{1}{2} \left ( e^{-(M-\alpha\omega) \tilde \sigma} - 
e^{-(M+\alpha \omega)\tilde \sigma} \right )
\label{sc1}
\end{equation}
From equation (\ref{sc1}) we can say that to get an exponentially decaying solution of $\tilde \sigma$ one must have $M \geq \alpha \omega$ which means $\beta^2 = \alpha^2 \omega^2 - M^2 < 0$. 
However, from equation (\ref{ansz}) one can see that if $\beta^2 < 0$ then $\phi(\tilde \tau, \tilde \sigma)$ will give rise to exponentially growing (or decaying) solutions in $\tilde \tau$. To get a 
solution which is oscillatory in 
$\tilde \tau$ or independent of $\tilde \tau$ one must have $\alpha \omega \geq M$, which immediately leads to a solution  growing exponentially with $\tilde \sigma$, if $\alpha \omega > M$. Thus, the only possibility to have a finite solution is $\beta=0$.

\noindent The perturbations ($\delta x^{i} = \phi^{\alpha} n^{i}_{\alpha}$) therefore,
turn out to be,
\begin{eqnarray}
\delta t &=& \phi(\tilde \tau, \tilde \sigma)n^0 = \epsilon~A e^{-\bar \sigma} \sinh \bar \sigma \, \, , \\
\delta \theta &=& \phi(\tilde \tau, \tilde \sigma)n^1 = -\epsilon~A e^{-\bar \sigma} \frac{C}{\omega}~\frac{\cosh \bar \sigma}{\sqrt{\cosh^2 \bar \sigma - \alpha^2}}\, \,, \\
\delta \phi &=& \phi(\tilde \tau,\tilde \sigma)n^2 = \epsilon~A e^{-\bar \sigma}
~\frac{\sinh \bar \sigma \cosh^2 \bar \sigma}{\cosh^2 \bar \sigma - \alpha^2}\,\, .
\end{eqnarray}

\begin{center}
\begin{figure}[h] 
 \begin{minipage}{14pc}
 \includegraphics[width=17pc]{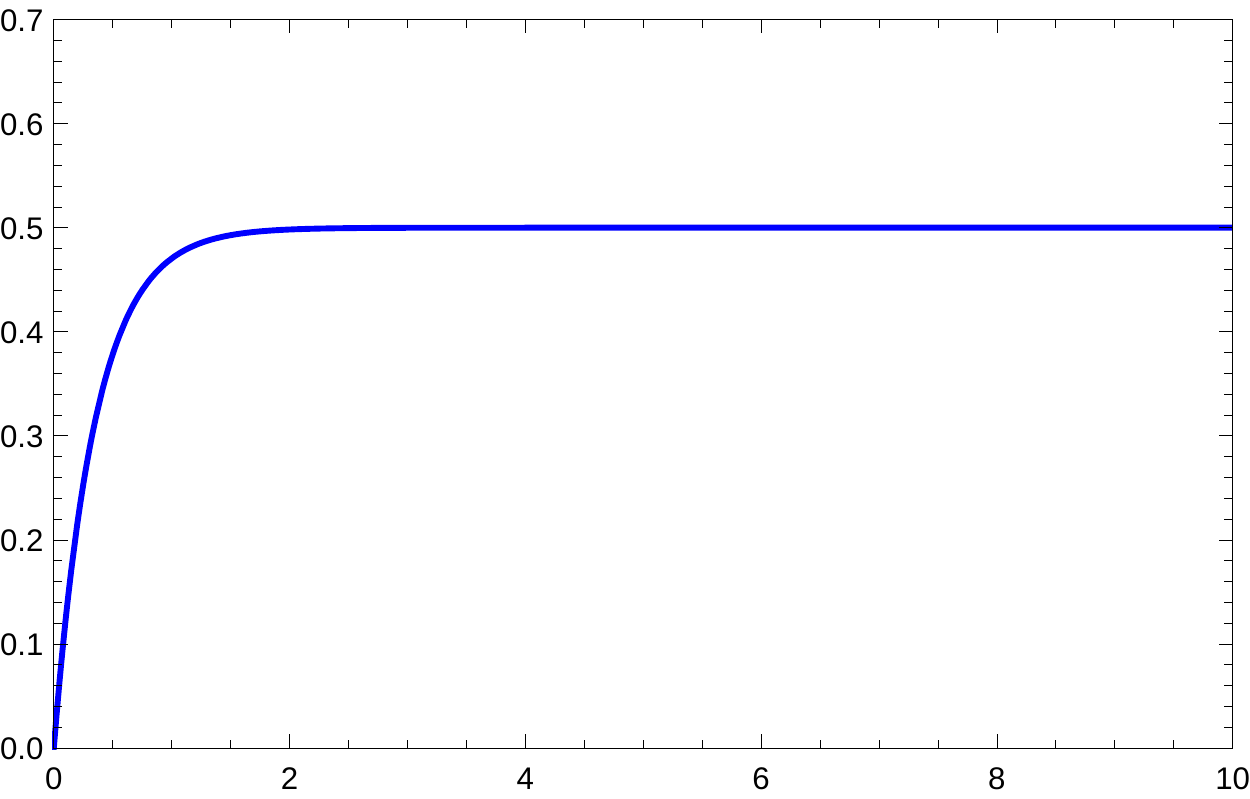}
 \put(0,-3){$\bf\tilde \sigma$}
 \put(-200,140){$ \delta t$}
 \caption{Plot of $\delta t$ for $A=1$, ~$\omega = \sqrt{3}$, ~$C=1$}
\label{dt}
\end{minipage}
\hspace{1.0in}
\begin{minipage}{14pc}
 \includegraphics[width=17pc]{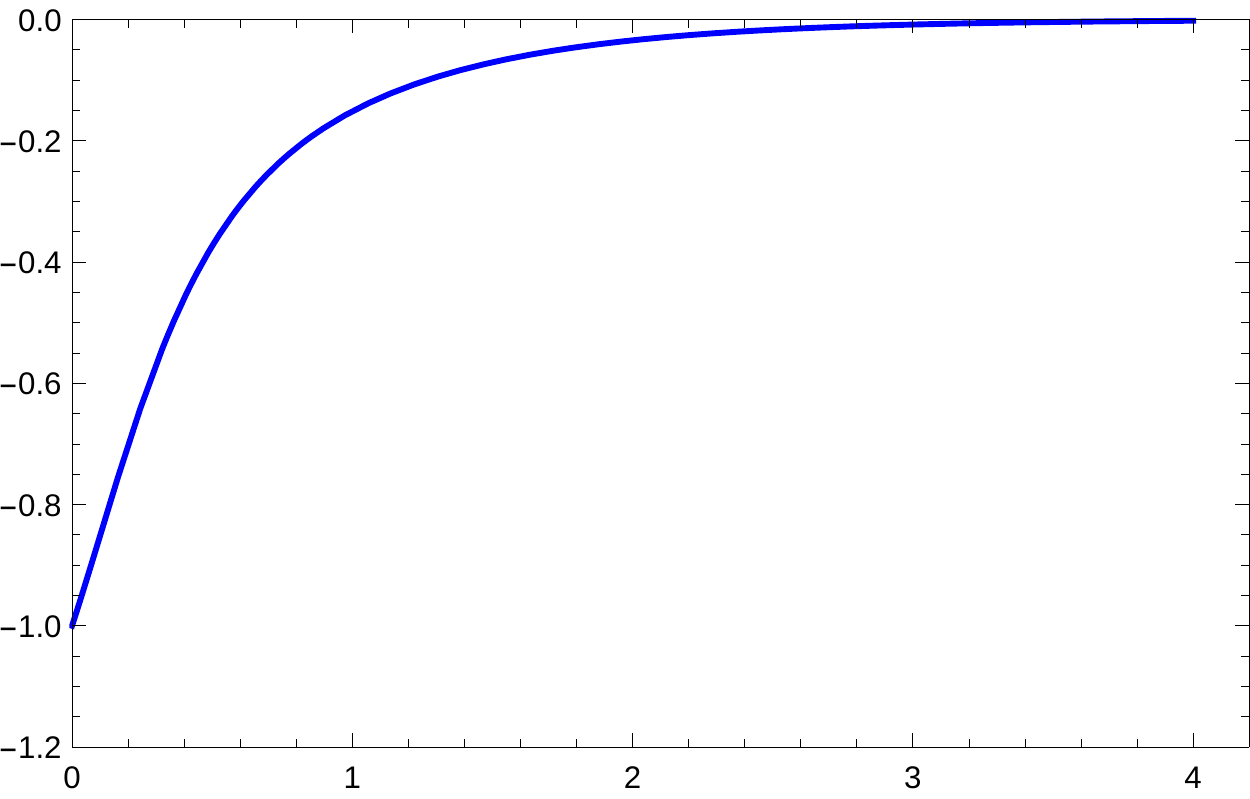}
 \put(0,-3){$\bf\tilde \sigma$}
 \put(-200,140){$ \delta \theta$}
 \caption{Plot of $\delta \theta$ for $A=1$, ~$\omega = \sqrt{3}$, ~$C=1$}
\label{dth}
\end{minipage}
\end{figure}
\end{center}

\begin{center}
\begin{figure}[h] 
 \begin{minipage}{12pc}
 \includegraphics[width=15pc]{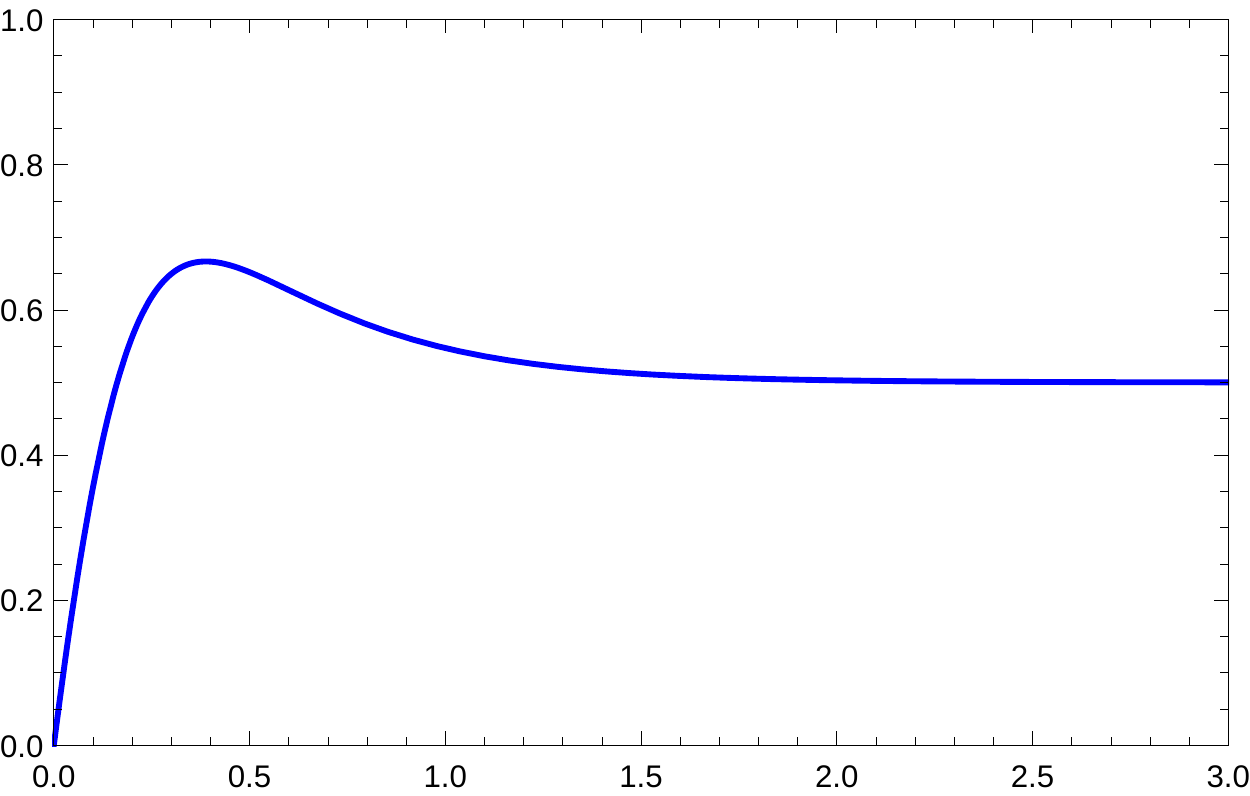}
 \put(0,-3){$\bf\tilde \sigma$}
 \put(-180,120){$ \delta \phi$}
 \caption{Plot of $\delta \phi$ for $A=1$, ~$\omega = \sqrt{3}$, ~$C=1$}
\label{dph}
\end{minipage}
\hspace{1.0in}
\begin{minipage}{12pc}
 \includegraphics[width=15pc]{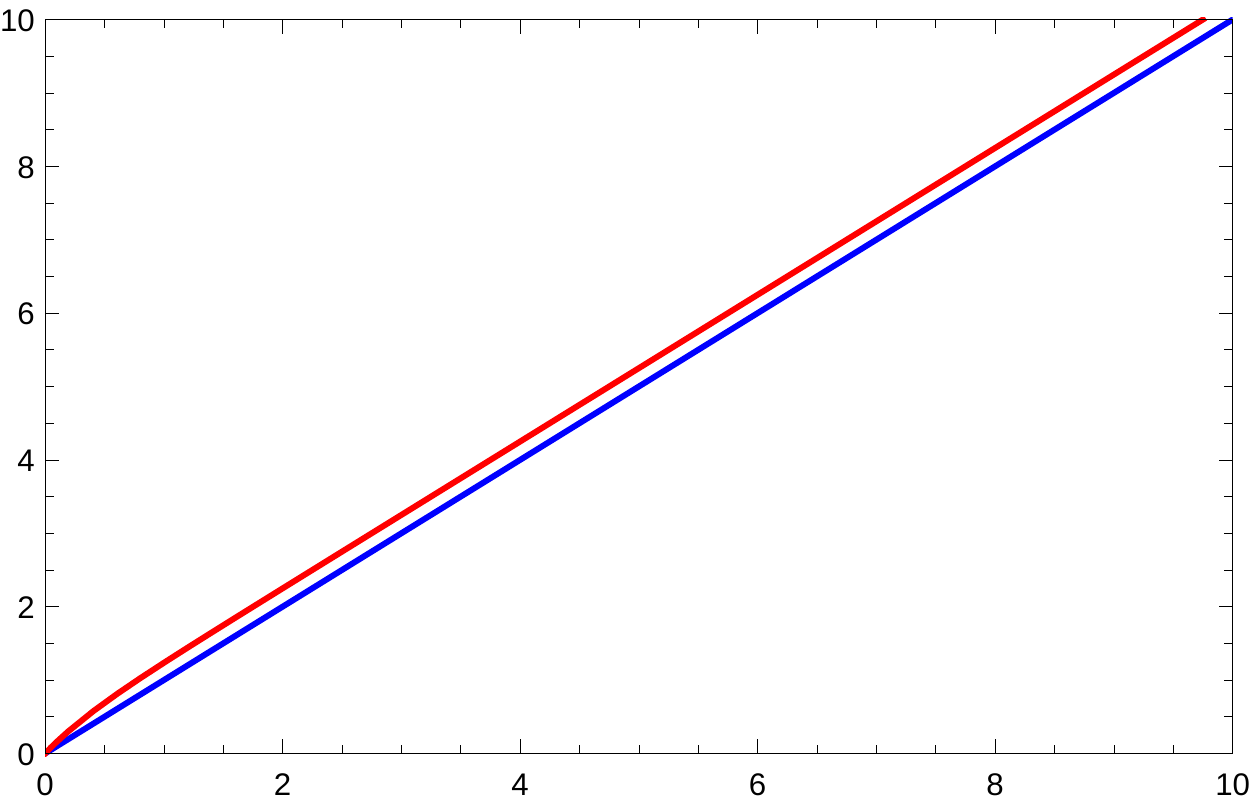}
 \put(0,-3){$\bf\tilde \sigma$}
 \put(-180,120){$t$}
 \caption{Plot of perturbed (red) and un-perturbed (blue) $t$- profile with $\tilde \tau = 0$ for $A=1$, ~$\omega = \sqrt{3}$, ~$C=1$ and $\epsilon = 0.5$}
\label{tp}
\end{minipage}
\end{figure}
\end{center}

\noindent All the perturbations are finite over the domain of $\tilde \sigma$. Note that there is no $\tilde \tau$ dependence since we have $\beta=0$. 
It is therefore only the zero mode of the perturbation which is
stable. All higher modes will basically decay or grow in time 
(imaginary $\beta$). Figures \ref{dt}, ~\ref{dth} and \ref{dph} have shown the perturbations  $\delta t$, ~$\delta \theta$ and ~$\delta \phi$ respectively for $A=1$, ~$\omega = \sqrt{3}$ and $C=1$. One can easily observe from the solutions and the figures that these perturbations are always finite. Hence the giant magnon is stable for the zero mode.
\noindent Let us now try to understand how this perturbation affects the
string profile. The perturbed embedding will be given as:
\begin{eqnarray}
t' &=& t + \delta t = \omega \tilde \tau + C\tilde \sigma +  \delta t \, \, , \\
\theta' &=& \theta + \delta \theta = \cos^{-1}\frac{\alpha}{\cosh \bar \sigma} + \delta \theta\, \, , \\
\phi' &=& \phi + \delta \phi = \omega\tilde \tau + g(\tilde \sigma)+ \delta \phi \, \, .
\end{eqnarray}
\begin{center}
\begin{figure}[h] 
 \begin{minipage}{12pc}
 \includegraphics[width=15pc]{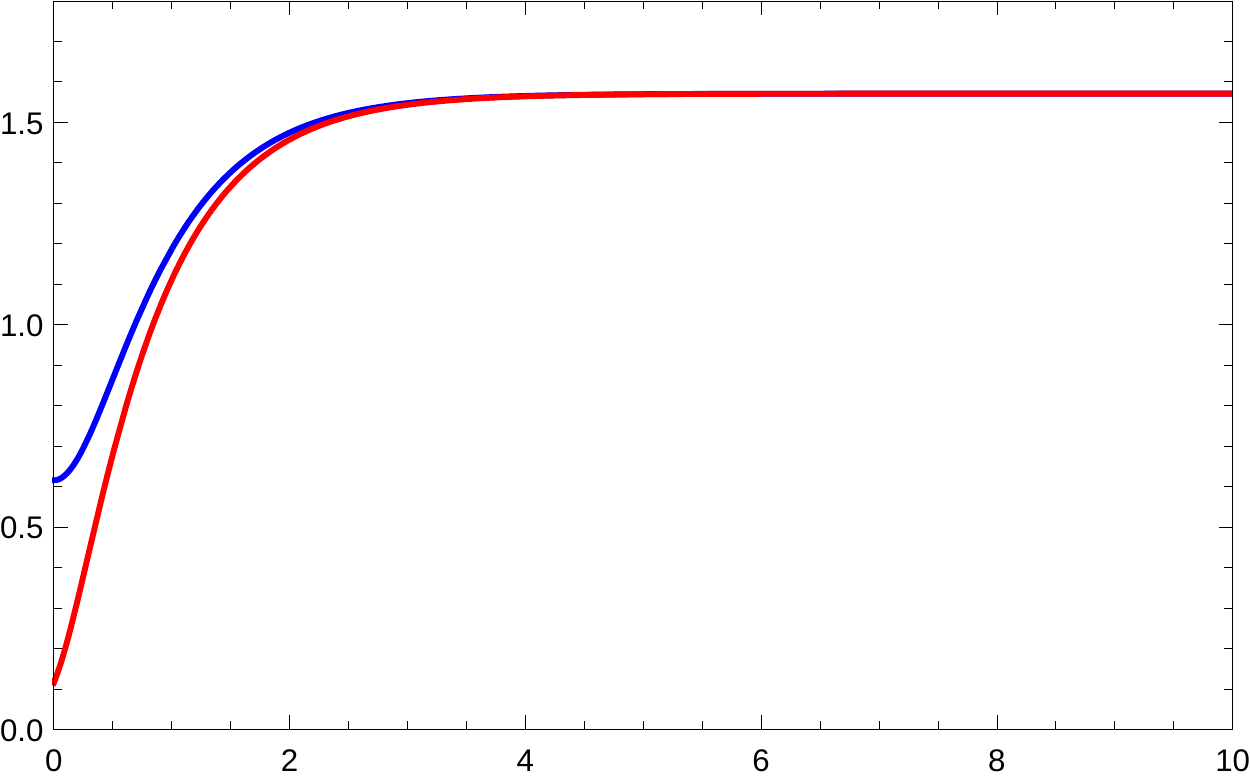}
 \put(0,-3){$\bf\tilde \sigma$}
 \put(-180,120){$\theta$}
 \caption{Plot of perturbed (red) and un-perturbed (blue) $\theta$-profile with $\tilde \tau = 0$ for $A=1$, ~$\omega = \sqrt{3}$, ~$C=1$ and $\epsilon = 0.5$}
\label{thp}
\end{minipage}
\hspace{1.0in}
\begin{minipage}{12pc}
 \includegraphics[width=15pc]{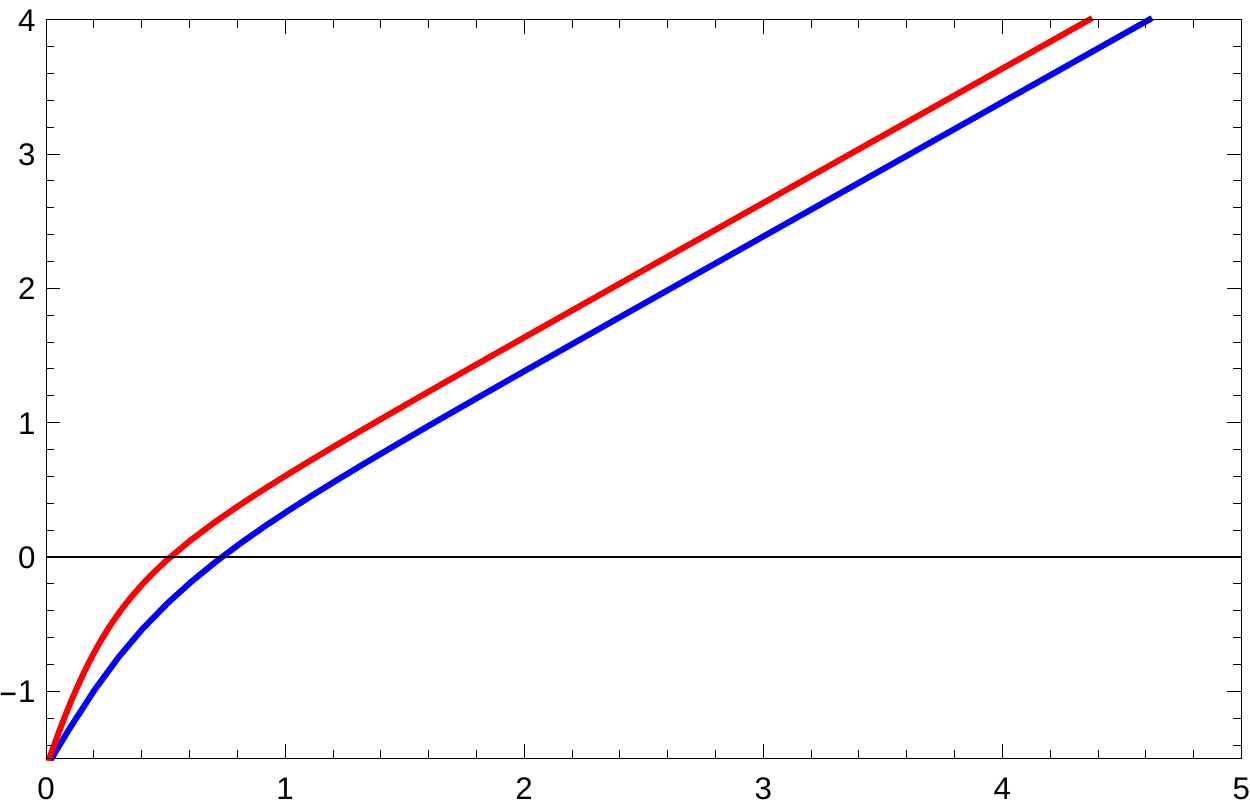}
 \put(0,-3){$\bf\tilde \sigma$}
 \put(-180,120){$\phi$}
 \caption{Plot of perturbed (red) and un-perturbed (blue) $\phi$-profile with $\tilde \tau = 0$ for $A=1$, ~$\omega = \sqrt{3}$, ~$C=1$ and $\epsilon = 0.5$}
\label{php}
\end{minipage}
\end{figure}
\end{center}

\noindent Figures \ref{tp}, ~\ref{thp} and \ref{php} describe the perturbed profiles for $t$, ~$\theta$ and $\phi$ respectively for $\tilde \tau = 0$. Here the blue curve describes the original un-perturbed profile whereas the red curve describes the perturbed profile. We have taken the amplitude of perturbation $\epsilon = 0.5$. One notices clearly that over the entire domain of $\tilde \sigma$, the perturbed profile is finite for finite $\tilde \sigma$, but different from the unperturbed one. Thus, the stability of the giant magnon is guaranteed through the stability of the zero
mode fluctuation, for all values of $\tilde \tau$.

\section{Single spike solution and perturbation in $\mathbb R \times S^2$}
\noindent In the single spike limit $|\omega| \rightarrow C$, from equation (\ref{vir1}) we have 
\begin{equation}
a= \frac{C^2}{\kappa} \, \, .
\end{equation} 
\noindent Equation (\ref{th1}) will take the following form
\begin{equation}
\theta' = \pm \frac{1}{\kappa \sin \theta}\sqrt{(\kappa^2 - C^2 \sin^2 \theta)(\kappa^2 \sin^2 \theta - C^2)} \, \, .
\label{th3}
\end{equation}
\noindent The equations of motion and the Virasoro constraints give us the following conditions,
\begin{eqnarray}
f'(\tilde \sigma) = a \hspace{0.2in}
; \hspace{0.2in} g'(\tilde \sigma) = \frac{C}{\sin^2 \theta} \, \, .
\end{eqnarray}
\noindent Integrating equation (\ref{th3}) we can write 
\begin{equation}
\sigma = \pm \frac{\kappa}{\sqrt{\kappa^2 - C^2}}~F \left(\sin^{-1}{\frac{\kappa \cos \theta}{\sqrt{\kappa^2 - C^2}},~-\frac{C^2}{\kappa^2}} \right) \, \, .
\end{equation}

\noindent Where $F$ is the elliptic integral of first kind. Now we can invert this expression by using the Jacobi elliptic functions. So the final form of $\theta$ will be 

\begin{equation}
\cos \theta = \frac{\sqrt{\kappa^2 - C^2}}{\kappa}~\sn(u,\tilde \kappa) \, \, .
\label{th4}
\end{equation}

\noindent Now in equation (\ref{th4}) the variable $u$ and the modulus $\tilde \kappa$ is defined  as 

\begin{eqnarray}
u = \pm \frac{\sqrt{\kappa^2 - C^2}}{\kappa} ~\tilde \sigma = \pm \sqrt{1 + \tilde \kappa} ~\tilde \sigma \hspace{0.2in} ;
\hspace{0.2in} \tilde \kappa = - \frac{C^2}{\kappa^2} \, \, .
\label{scn}
\end{eqnarray}
\subsection{Perturbations and stability of single spike in $\mathbb R \times S^2$}

\noindent The induced metric is given as:
\begin{equation}
ds^2 = (\kappa^2 - C^2 \sin^2 \theta) \left( -d{\tilde\tau}^2 + d{\tilde \sigma}^2 \right) \, \, .
\end{equation}

\noindent The tangent vectors to the worldsheet are given as:
\begin{equation}
e^{i}_{\tilde\tau} = \left (\kappa, ~0, ~C\right ) \hspace{0.2in}; \hspace{0.2in}
e^{i}_{\tilde \sigma} = \left ( f', ~\theta', ~g'\right ) \, \, .
\end{equation}

\noindent The normal to the worldsheet is given as:
\begin{equation}
n^{i} = \left( \frac{C}{\kappa}\frac{\sqrt{\kappa^2 \sin^2 \theta - C^2}}{\sqrt{\kappa^2 - C^2 \sin^2 \theta}}, ~-\frac{C}{\kappa \sin \theta}, ~\frac{\sqrt{\kappa^2 \sin^2 \theta - C^2}}{\sin^2 \theta \sqrt{\kappa^2 - C^2 \sin^2 \theta}} \right)
\label{nrm2}
\end{equation}
\noindent The extrinsic curvature tensor which is defined as $K_{ab}^{(\alpha)} = -g_{ij}( e^k_{a}\nabla_k e^i_b) n^{j(\alpha)}$, turns out to be
\begin{equation}
K_{ab} = \begin{pmatrix} -\frac{C^3}{\kappa} \cos \theta & -C\kappa \cos \theta \cr -C\kappa \cos \theta & -\frac{C^3}{\kappa} \cos \theta \end{pmatrix} .
\end{equation}

\noindent Using all of the above-stated quantities which appear
in the perturbation equation and after some lengthy algebra, 
one arrives at the
following simple equation for the perturbation scalar $\phi$
\begin{equation}
\left( -\partial_{\tilde\tau}^2 + \partial_{\tilde\sigma}^2 \right) \phi + \left(\frac{(\kappa^4 - C^4)(\kappa^2 - C^2 -C^2 \sin^2 \theta)}{
\kappa^2(\kappa^2 - C^2 + C^2 \cos^2 \theta)}\right) \phi = 0 \, \, .
\label{pert2} 
\end{equation}

\noindent Let us now use the same ansatz as before
\begin{equation}
\phi(\tilde \tau, \tilde \sigma) = \epsilon e^{i \beta \tilde\tau}~P(\tilde\sigma) \, \, ,
\label{fs1}
\end{equation}
\noindent where $\beta$ is the eigenvalue and $\epsilon$, a constant which we may relate to the
amplitude of the perturbation. It must be emphasized that $\epsilon$
has to be small in value (i.e. $\epsilon<<1$) in order to ensure that the
deformation is genuinely a small perturbation.

\noindent Using the previous ansatz to separate the $\tilde \tau$ and $\tilde \sigma$ part and exploiting equation (\ref{th4}) we can write down the equation for $P(\tilde \sigma)$ as

\begin{equation}
\frac{d^2 P}{du^2} + \left( \tilde \beta^2 + \frac{(1-\tilde \kappa)
(1 + \tilde \kappa ~\sn^2 u)}{(1 + \tilde \kappa) (1 - \tilde \kappa ~\sn^2 u)} \right) P =  0
\label{pert3}
\end{equation}

\noindent Where $\tilde \beta$ is defined as 
$\tilde \beta^2 = \frac{\beta^2 \kappa^2}{\kappa^2 - C^2} = \frac{\beta^2}{1+\tilde \kappa}$. 

\noindent We can examine the nature of the potential $V(u) = \frac{(1-\tilde \kappa)
(1 + \tilde \kappa ~\sn^2 u)}{(1 + \tilde \kappa) (1 - \tilde \kappa ~\sn^2 u)}$ graphically by plotting it explicitly. Before that some points need to be mentioned: 1) Since $\tilde \sigma$ runs from $0$ to $\pm \infty$ so is $u$ except at $\tilde \kappa = -1$ where $u=0$ , one can see that from eqn (\ref{scn}), which is clearly justified from the condition of Lorenzian signature of the induced metric, 2) From (\ref{scn}) one can see that $-1<\tilde \kappa \leq 0$. 

\begin{center}
\begin{figure}[h] 
 \includegraphics[width=30pc]{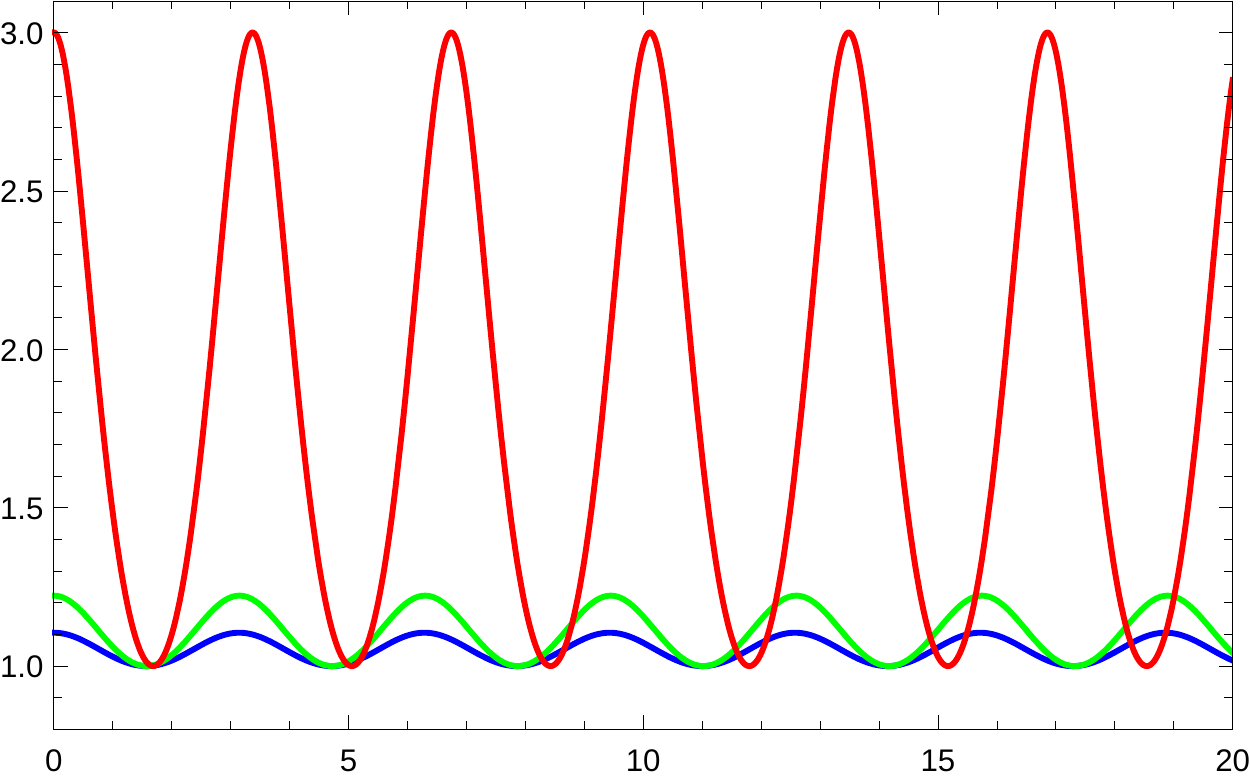}
 \put(0,-3){$\bf u$}
 \put(-390,180){$\bf  V(u)$}
 \caption{Plot of $V(u)$ for three different values of $\tilde \kappa$: 1) $\tilde \kappa = -0.05$ (blue), 2) $\tilde \kappa = -0.1$ (green) \& 3) $\tilde \kappa = -0.5$ (red)}
\label{pot1}
\end{figure}
\end{center}

\noindent Plots of $V(u)$ have been shown in figures (\ref{pot1}). Figure (\ref{pot1}) shows the nature of the potential $V(u)$ in general for three different values of $\tilde \kappa$: 1) $\tilde \kappa = -0.05$ which is shown by the blue curve, 2)$\tilde \kappa = -0.1$ which is shown by the green curve and 3) $\tilde \kappa = -0.5$ which is shown by the red curve. From all these three curves one can see that the potential is of periodic nature. 

\subsection{Solution of the perturbation equation}\label{subsol}
We have solved the perturbation equation (\ref{pert3}) numerically using the finite difference method. We transform the above differential equation into a set of algebraic equations represented in the form of a matrix. We compute the eigenvalues and eigen-vectors for different cases. Here we consider mainly three different cases: 1) $\tilde \kappa = -0.05$, 2) $\tilde \kappa = -0.1$ and 3) $\tilde \kappa = -0.5$.  The corresponding potential functions are shown in Figure 7
(the blue, green and red curves respectively).  

\begin{table}
\caption{Eigenvalues for $\tilde \kappa = -0.05$}\label{TT1}
\begin{center}
\begin{tabular}{|c|c|c|c|c|c|c|} \hline 
  $\tilde \kappa$ & N & $\lambda_0$ & $\lambda_1$ & $\lambda_2$ & $\lambda_3$ & $\lambda_4$ \\
  \hline
  & $150$ &  $-1.05231984$ & $2.94240006$ & $2.94271174$ & $14.91869654$ & $14.91869660$\\
    & $200$ & $-1.05231981$ & $2.94265559$ & $2.94296730$ & $14.92278399$ & $14.92278406$ \\
    & $250$ & $-1.05231980$ & $2.94277387$ & $2.94308559$ & $14.92467619$ & $14.92467625$  \\
   
  $-0.05$ & $300 $ & $-1.05231980$ & $2.94283812$ & $2.94314984$ & $14.92570412$ & $14.92570418$  \\
   & $350$ & $-1.05231979$ & $2.94287686$ & $2.94318859$ & $14.92632395$ & $14.92632401$ \\
   
   & $400$ & $-1.05231979$ & $2.94290201$ & $2.94321374$ & $14.92672626$ & $14.92672632$ \\
   
   & $450$ & $-1.05231979$ & $2.94291925$ & $2.94323098$ & $14.92700209$ & $14.92700215$ \\
   
   & $500$ & $-1.05231979$ & $2.94293158$ & $2.94324331$ & $14.92719939$ & $14.92719945 $ \\
  
\hline
 
\end{tabular}
\label{T1}
\end{center}
\end{table}

\begin{table}
\caption{Eigenvalues for $\tilde \kappa = -0.1$}\label{TT2}
\begin{center}
\begin{tabular}{|c|c|c|c|c|c|c|} \hline 
  $\tilde \kappa$ & N & $\lambda_0$ & $\lambda_1$ & $\lambda_2$ & $\lambda_3$ & $\lambda_4$ \\
  \hline
  & $150$ &  $-1.10987439$ & $2.87095432$ & $2.87219004$ & $14.80231858$ & $14.80231957$\\
    & $200$ & $-1.10987429$ & $2.87120887$ & $2.87244469$ & $14.80639068$ & $14.80639168$ \\
    & $250$ & $-1.10987425$ & $2.87132670$ & $2.87256257$ & $14.80827577$ & $14.80827677$  \\
   
  $-0.1$ & $300 $ & $-1.10987422$ & $2.87139071$ & $2.87262660$ & $14.80929984$ & $14.80930084$  \\
   & $350$ & $-1.10987421$ & $2.87142930$ & $2.87266521$ & $14.80991735$ & $14.80991835$ \\
   
   & $400$ & $-1.10987420$ & $2.87145435$ & $2.87269027$ & $14.81031815$ & $14.81031915$ \\
   
   & $450$ & $-1.10987419 $ & $2.87147153$ & $2.87270745$ & $14.81059294$ & $14.81059394$ \\
   
   & $500$ & $-1.10987418$ & $2.87148381$ & $2.87271974$ & $14.81078949$ & $14.81079049 $ \\
  
\hline
 
\end{tabular}
\label{T2}
\end{center}
\end{table}

\begin{table}
\caption{Eigenvalues for $\tilde \kappa = -0.5$}\label{TT3}
\begin{center}
\begin{tabular}{|c|c|c|c|c|c|c|} \hline 
  $\tilde \kappa$ & N & $\lambda_0$ & $\lambda_1$ & $\lambda_2$ & $\lambda_3$ & $\lambda_4$ \\
  \hline
  & $150$ &  $-2.00227136$ & $1.65322654$ & $1.65625830$ & $12.02953349$ & $12.03027567$\\
    & $200$ & $-2.00226233$ & $1.65345610$ & $1.65647916$ & $12.03309099$ & $12.03383350$ \\
    & $250$ & $-2.00225814$ & $1.65356236$ & $1.65658139$ & $12.03473785$ & $12.03548050$  \\
   
  $-0.5$ & $300 $ & $- 2.00225587$ & $1.65362008$ & $1.65663693$ & $12.03563250$ & $12.03637524$  \\
   & $350$ & $-2.00225450$ & $1.65365488$ & $1.65667041$ & $12.03617197$ & $12.03691476$ \\
   
   & $400$ & $-2.00225361$ & $1.65367747$ & $1.65669215$ & $12.03652211$ & $12.03726494$ \\
   
   & $450$ & $-2.00225300 $ & $1.65369296$ & $1.65670705$ & $12.03676218$ & $12.03750502$ \\
   
   & $500$ & $-2.00225256$ & $1.65370404$ & $1.65671770$ & $12.03693389$ & $12.03767675 $ \\
  
\hline
 
\end{tabular}
\label{T3}
\end{center}
\end{table}

\begin{table}
\caption{$\Delta \lambda$ for $\tilde \kappa = -0.05$}\label{TT4}
\begin{center}
\begin{tabular}{|c|c|c|c|c|c|} \hline 
   $\Delta \lambda$  & $\Delta \lambda_0$ & $\Delta \lambda_1 $ & $\Delta \lambda_2 $ & $ \Delta \lambda_3 $ & $\Delta \lambda_4 $ \\
  \hline
   $|\lambda(200) - \lambda(150)|$ &  $3.0 \times 10^{-8}$ & $2.55 \times 10^{-4}$ & $2.55 \times 10^{-4}$ & $4.1 \times 10^{-3}$ & $4.1 \times 10^{-3}$\\
     $|\lambda(250) - \lambda(200)|$ & $1.0 \times 10^{-8}$ & $1.2 \times 10^{-4}$ & $1.2 \times 10^{-4}$ & $1.9 \times 10^{-3}$ & $1.9\times 10^{-3}$ \\
     $|\lambda(300) - \lambda(250)|$ & $0.0 \times 10^{-8}$ & $0.6 \times 10^{-4}$ & $0.64 \times 10^{-4 }$ & $1.0 \times 10^{-3}$ & $1.0 \times 10^{-3}$  \\
   
   $|\lambda(350) - \lambda(300)| $ & $0.0 \times 10^{-8}$ & $0.30 \times 10^{-4}$ & $0.38 \times 10^{-4}$ & $0.6 \times 10^{-3}$ & $0.62 \times 10^{-3}$  \\
    $|\lambda(400) - \lambda(350)|$ & $0.0 \times 10^{-8}$ & $0.2 \times 10^{-4}$ & $0.25 \times 10^{-4}$ & $0.4 \times 10^{-3}$ & $0.4 \times 10^{-3}$ \\
   
    $|\lambda(450) - \lambda(400)|$ & $0.0 \times 10^{-8}$ & $0.17 \times 10^{-4}$ & $0.17 \times 10^{-4}$ & $0.27 \times 10^{-3}$ & $0.27 \times 10^{-3}$ \\
   
    $|\lambda(500) - \lambda(450)|$ & $0.0 \times 10^{-8}$ & $0.12 \times 10^{-4}$ & $0.12 \times 10^{-4}$ & $0.19 \times 10^{-3}$ & $0.19 \times 10^{-3}$ \\

\hline
 
\end{tabular}
\label{T4}
\end{center}
\end{table}

\begin{table}
\caption{$\Delta \lambda$ for $\tilde \kappa = -0.1$}\label{TT5}
\begin{center}
\begin{tabular}{|c|c|c|c|c|c|} \hline 
   $\Delta \lambda$  & $\Delta \lambda_0$ & $\Delta \lambda_1 $ & $\Delta \lambda_2 $ & $ \Delta \lambda_3 $ & $\Delta \lambda_4 $ \\
  \hline
   $|\lambda(200) - \lambda(150)|$ &  $1.0 \times 10^{-7}$ & $2.5 \times 10^{-4}$ & $2.54 \times 10^{-4}$ & $4.1 \times 10^{-3}$ & $4.1 \times 10^{-3}$\\
     $|\lambda(250) - \lambda(200)|$ & $0.4 \times 10^{-7}$ & $1.17 \times 10^{-4}$ & $1.2 \times 10^{-4}$ & $1.9 \times 10^{-3}$ & $1.9\times 10^{-3}$ \\
     $|\lambda(300) - \lambda(250)|$ & $0.3 \times 10^{-7}$ & $0.64 \times 10^{-4}$ & $0.64 \times 10^{-4 }$ & $1.0 \times 10^{-3}$ & $1.0 \times 10^{-3}$  \\
   
   $|\lambda(350) - \lambda(300)| $ & $0.1 \times 10^{-7}$ & $0.38 \times 10^{-4}$ & $0.38 \times 10^{-4}$ & $0.6 \times 10^{-3}$ & $0.61 \times 10^{-3}$  \\
    $|\lambda(400) - \lambda(350)|$ & $0.1 \times 10^{-7}$ & $0.25 \times 10^{-4}$ & $0.25 \times 10^{-4}$ & $0.4 \times 10^{-3}$ & $0.4 \times 10^{-3}$ \\
   
    $|\lambda(450) - \lambda(400)|$ & $0.1 \times 10^{-7}$ & $0.17 \times 10^{-4}$ & $0.17 \times 10^{-4}$ & $0.27 \times 10^{-3}$ & $0.27 \times 10^{-3}$ \\
   
    $|\lambda(500) - \lambda(450)|$ & $0.1 \times 10^{-7}$ & $0.12 \times 10^{-4}$ & $0.12 \times 10^{-4}$ & $0.19 \times 10^{-3}$ & $0.2 \times 10^{-3}$ \\

\hline
 
\end{tabular}
\label{T5}
\end{center}
\end{table}

\begin{table}
\caption{$\Delta \lambda$ for $\tilde \kappa = -0.5$}\label{TT6}
\begin{center}
\begin{tabular}{|c|c|c|c|c|c|} \hline 
   $\Delta \lambda$  & $\Delta \lambda_0$ & $\Delta \lambda_1 $ & $\Delta \lambda_2 $ & $ \Delta \lambda_3 $ & $\Delta \lambda_4 $ \\
  \hline
   $|\lambda(200) - \lambda(150)|$ &  $9.0 \times 10^{-6}$ & $2.3 \times 10^{-4}$ & $2.2 \times 10^{-4}$ & $3.6 \times 10^{-3}$ & $3.6 \times 10^{-3}$\\
     $|\lambda(250) - \lambda(200)|$ & $4.2 \times 10^{-6}$ & $1.0 \times 10^{-4}$ & $1.0 \times 10^{-4}$ & $1.6 \times 10^{-3}$ & $1.65\times 10^{-3}$ \\
     $|\lambda(300) - \lambda(250)|$ & $2.27 \times 10^{-6}$ & $0.6 \times 10^{-4}$ & $0.6 \times 10^{-4 }$ & $0.9 \times 10^{-3}$ & $0.9 \times 10^{-3}$  \\
   
   $|\lambda(350) - \lambda(300)| $ & $1.37 \times 10^{-6}$ & $0.35 \times 10^{-4}$ & $0.33 \times 10^{-4}$ & $0.54 \times 10^{-3}$ & $0.54 \times 10^{-3}$  \\
    $|\lambda(400) - \lambda(350)|$ & $0.9 \times 10^{-6}$ & $0.22 \times 10^{-4}$ & $0.22 \times 10^{-4}$ & $0.35 \times 10^{-3}$ & $0.35 \times 10^{-3}$ \\
   
    $|\lambda(450) - \lambda(400)|$ & $0.6 \times 10^{-6}$ & $0.15 \times 10^{-4}$ & $0.14 \times 10^{-4}$ & $0.24 \times 10^{-3}$ & $0.24 \times 10^{-3}$ \\
   
    $|\lambda(500) - \lambda(450)|$ & $0.44 \times 10^{-6}$ & $0.11 \times 10^{-4}$ & $0.1 \times 10^{-4}$ & $0.17 \times 10^{-3}$ & $0.17 \times 10^{-3}$ \\

\hline
 
\end{tabular}
\label{T6}
\end{center}
\end{table}
\noindent The first few eigenvalues for this three different cases are listed below. The eigenvalues are denoted by $\lambda$ and $\lambda = \tilde \beta^2$ and $N$ denotes the number of points we have taken during the computation. One can see from the above tables of eigenvalues that as we increase the number of points, the eigenvalues are converging. To understand it  in a better way, we can construct a quantity like  $\Delta \lambda = |\lambda_0(200)- \lambda_0(150)|$ i.e. taking differences between
two subsequent rows corresponding to different number of points for same set of eigenvalues. We have shown these differences in table (\ref{T4}), (\ref{T5}) and (\ref{T6}). We can see from the tables that as we increase the number of points, difference between the eigenvalues corresponding to subsequent number of points will decrease accordingly. 
\begin{center}
\begin{figure}[h] 
 \begin{minipage}{10pc}
 \includegraphics[width=10pc]{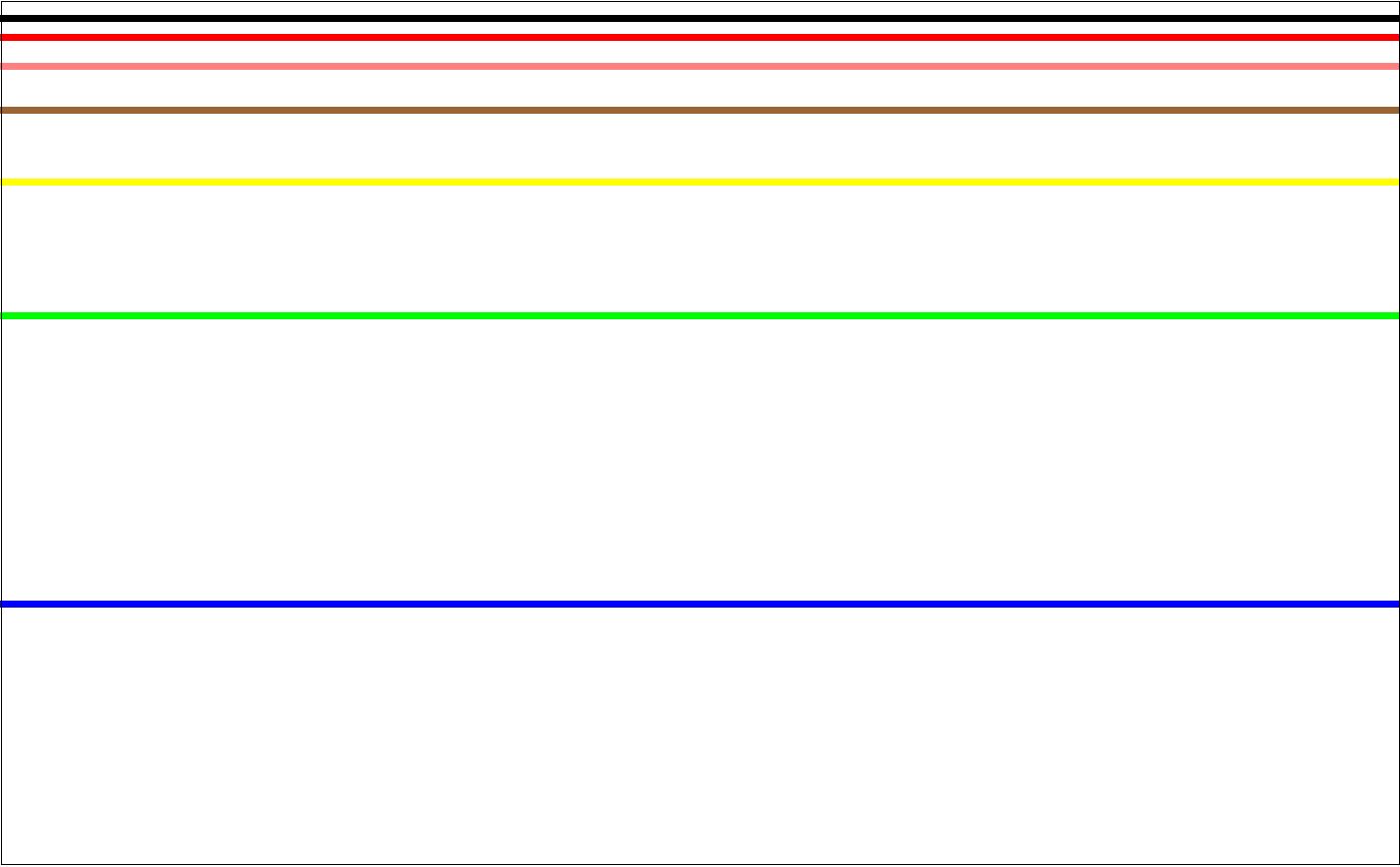}
 \put(-85,-10){$\tilde \kappa = -0.05$}
\label{l1}
\end{minipage}
\hspace{0.2in}
\begin{minipage}{10pc}
 \includegraphics[width=10pc]{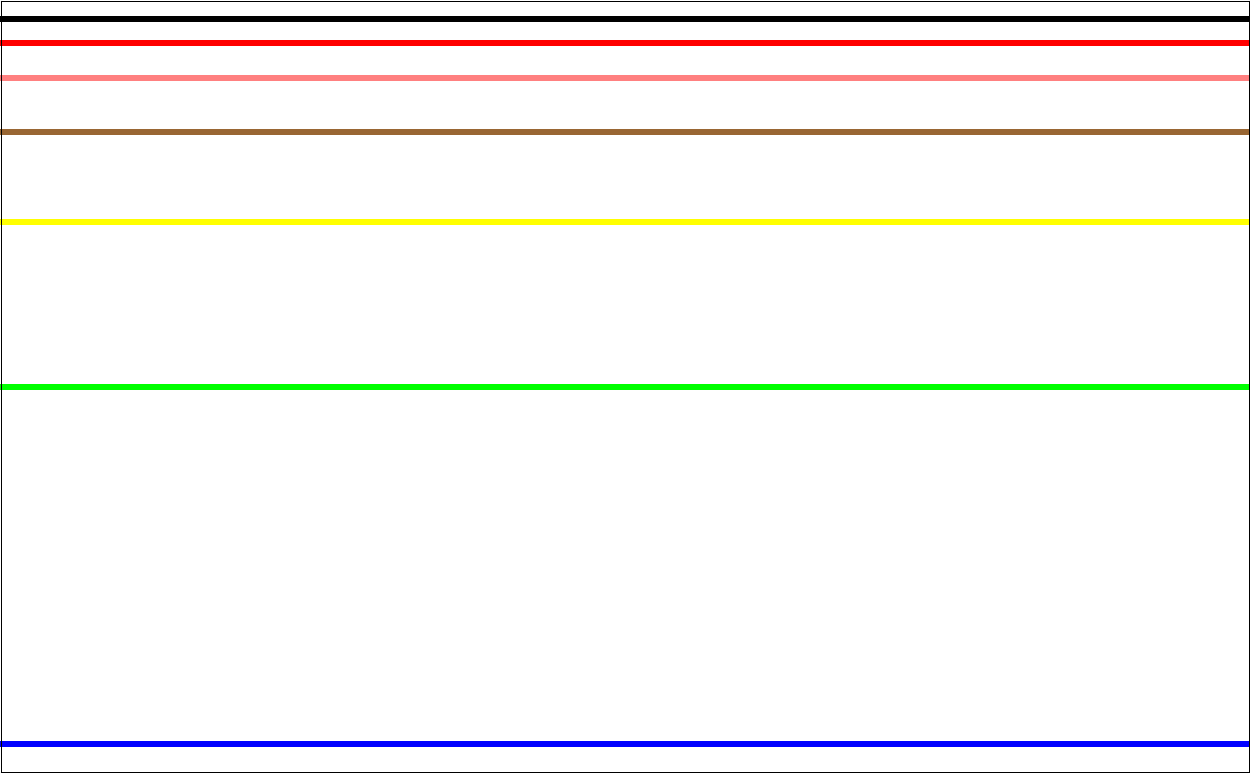}
 \put(-85,-10){$ \tilde \kappa = -0.1$}
 \label{l2}
\end{minipage}
\hspace{0.2in}
\begin{minipage}{10pc}
 \includegraphics[width=10pc]{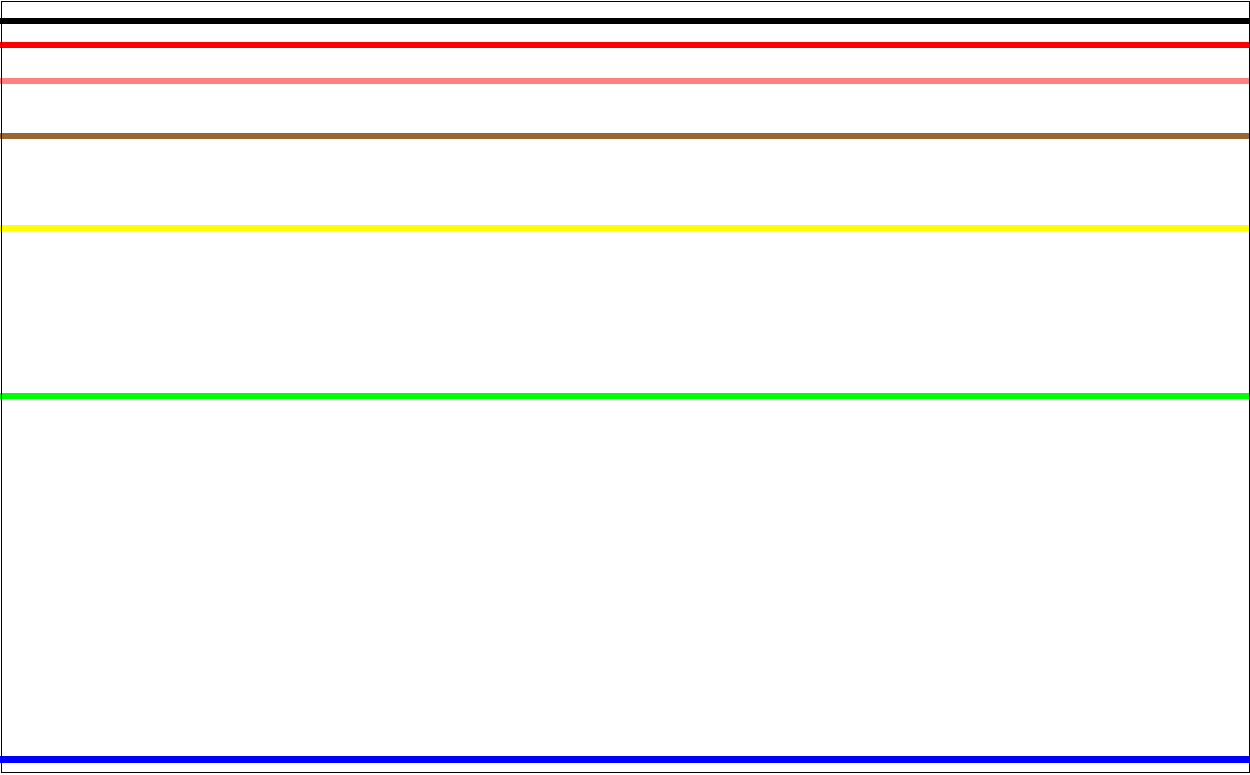}
 \put(-85,-10){$\tilde \kappa = -0.5$}
\label{l3}
\end{minipage}
\caption{Eigenvalue $\lambda_4$ is shown for three different cases of $\tilde \kappa$ where different colored lines correspond $N=150$ (blue), $N=200$ (green), $N=250$ (yellow), $N=300$ (brown), $N=350$ (pink), $N=400$ (red) and $N=450$ (black) }
\label{l1}
\end{figure}
\end{center}
\noindent To get a clear notion about this fact we have  shown, in Figure 8, the eigenvalue $\lambda_4$ for three cases $\tilde \kappa = -0.05$, $\tilde \kappa = -0.1$ and $\tilde \kappa = -0.5$. 
\begin{center}
\begin{figure}[h] 
 \begin{minipage}{20pc}
 \includegraphics[width=20pc]{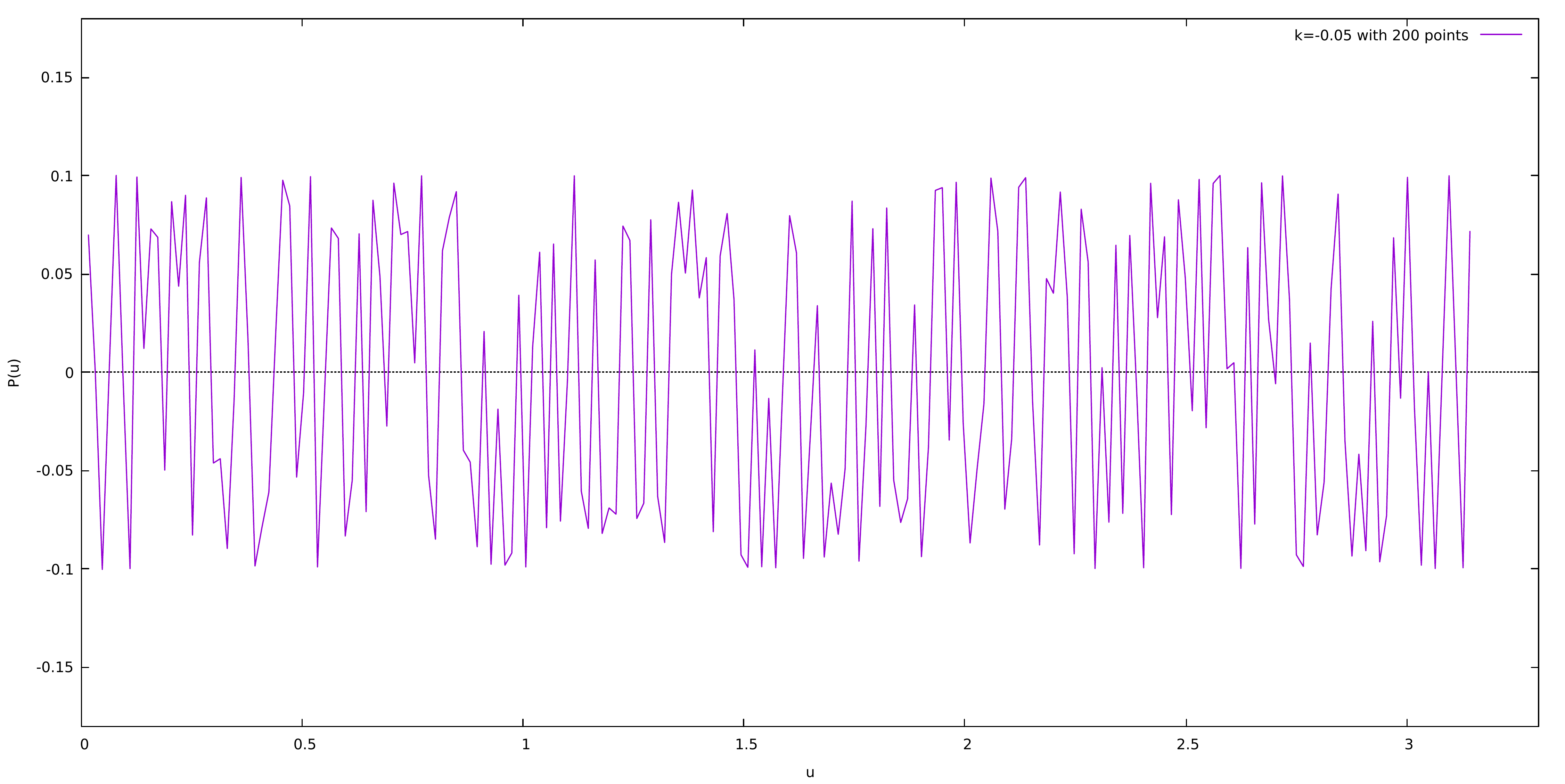}
 \put(-85,-10){$ \bf u$}
 \put(-250,130){$\bf  P(u)$}
 \caption{$P(u)$ vs $u$ plot corresponding to $\lambda_0$ with $\tilde \kappa = -0.05$ for $N=200$ points}
\label{sp1}
\end{minipage}
\hspace{0.1in}
\begin{minipage}{20pc}
 \includegraphics[width=20pc]{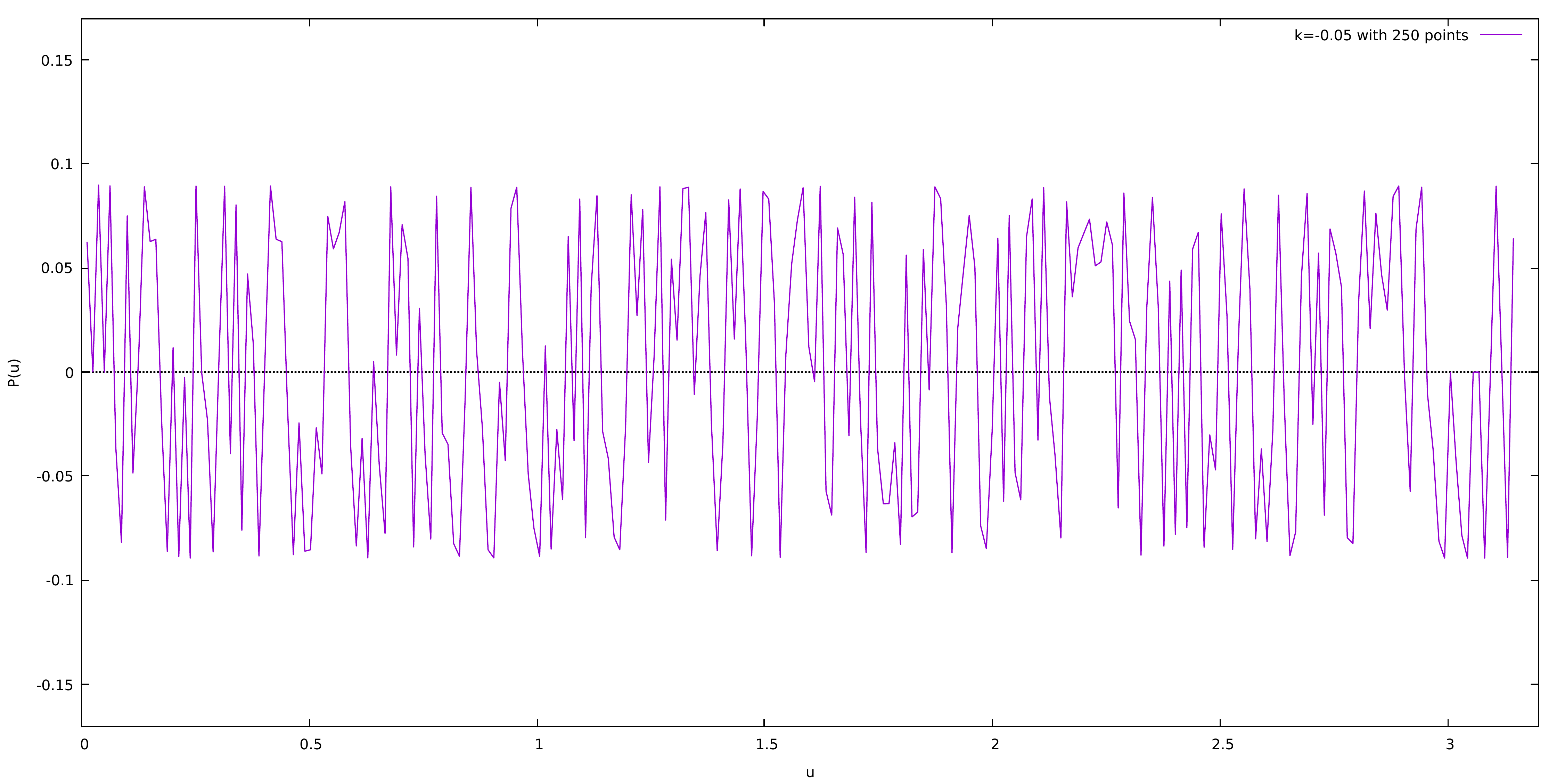}
 \put(-85,-10){$\bf u$}
 \put(-250,130){$\bf P(u)$}
 \caption{$P(u)$ vs $u$ plot corresponding to $\lambda_0$ with $\tilde \kappa = -0.05$ for $N=250$ points}
 \label{sp2}
\end{minipage}
\end{figure}
\end{center}

\begin{center}
\begin{figure}[h] 
 \begin{minipage}{20pc}
 \includegraphics[width=20pc]{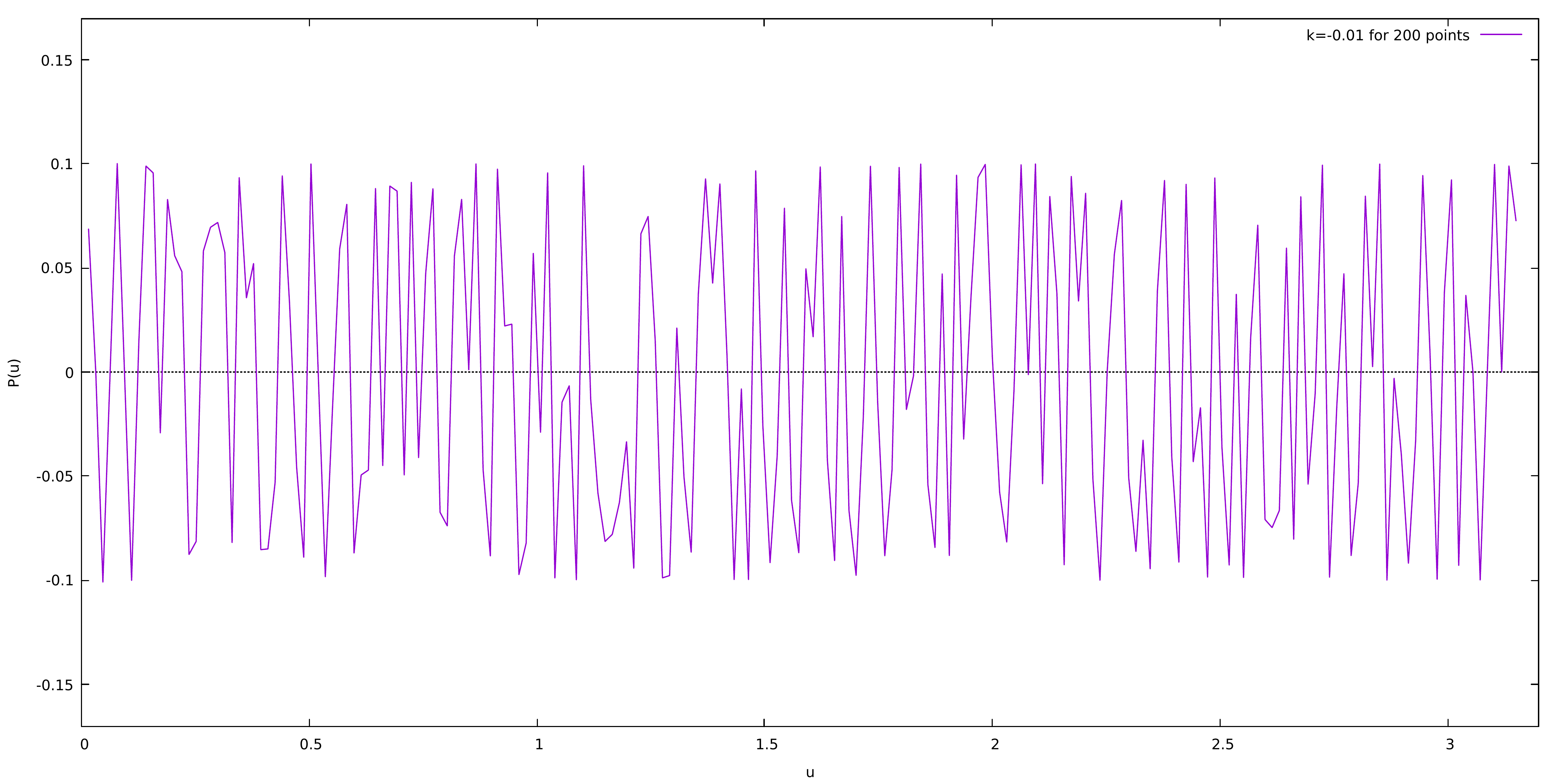}
 \put(-85,-10){$ \bf u$}
 \put(-250,130){$\bf  P(u)$}
 \caption{$P(u)$ vs $u$ plot corresponding to $\lambda_0$ with $\tilde \kappa = -0.1$ for $N=200$ points}
\label{sp3}
\end{minipage}
\hspace{0.1in}
\begin{minipage}{20pc}
 \includegraphics[width=20pc]{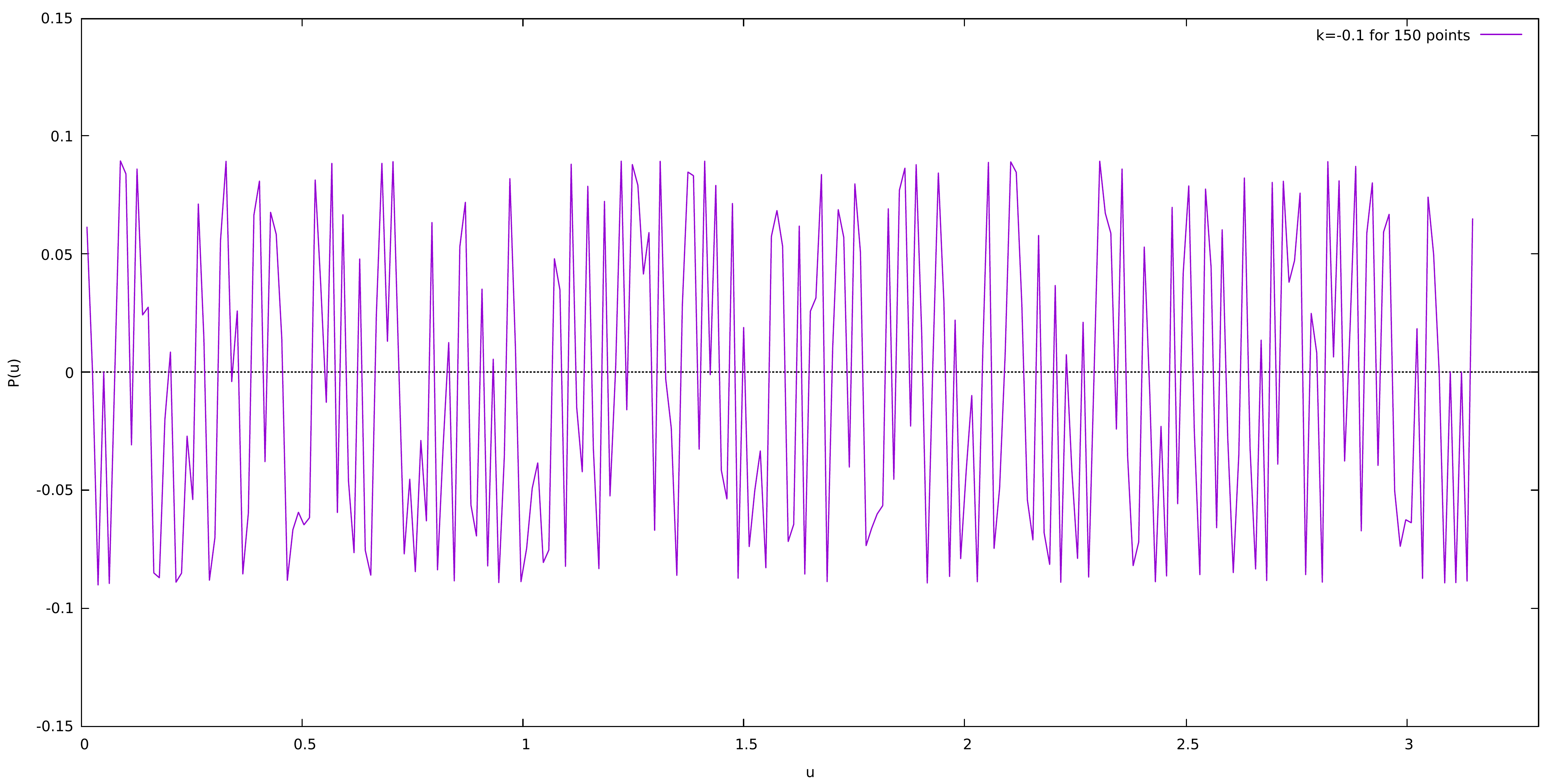}
 \put(-85,-10){$\bf u$}
 \put(-250,130){$\bf P(u)$}
 \caption{$P(u)$ vs $u$ plot corresponding to $\lambda_0$ with $\tilde \kappa = -0.1$ for $N=250$ points}
 \label{sp4}
\end{minipage}
\end{figure}
\end{center}

\noindent In each of these cases we plot $\lambda_4$ for different number of points. Each horizontal coloured line represents $\lambda_4$ corresponding to $N = 150, 200 ,...$ etc. For all three cases colored lines correspond $N=150$ (blue), $N=200$ (green), $N=250$ (yellow), $N=300$ (brown), $N=350$ (pink), $N=400$ (red) and $N=450$ (black).
From this plot we can now clearly understand that as we increase number of points, difference between $\lambda$ (which we have denoted as $\delta \lambda$) will decrease gradually i.e in the above diagram the blue line ($\lambda_4$ corresponding to $N=150$) is far apart from the green line ($\lambda_4$ corresponding to $N = 200$). However the pink line ($N=350$) and red line ($N=400$) almost coincide and hence proves the idea of convergence. If we look at the lower eigenvalues (like $\lambda_0$, $\lambda_1$, $\lambda_2$, $\lambda_3$) we can see that the convergence is far better for them (clearly  visible from table (\ref{T4}), (\ref{T5}) and (\ref{T6}).

\begin{center}
\begin{figure}[h] 
 \begin{minipage}{20pc}
 \includegraphics[width=20pc]{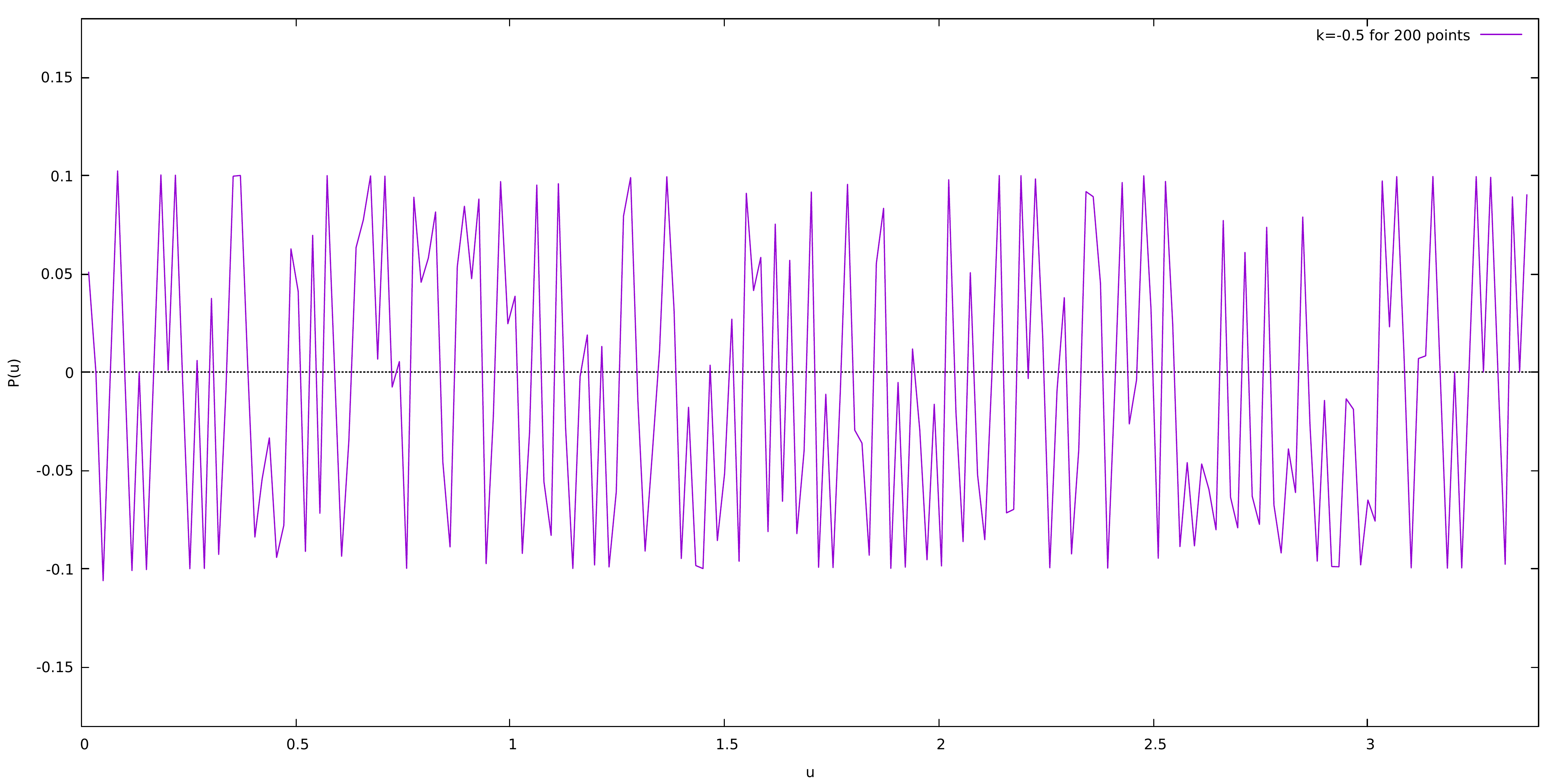}
 \put(-85,-10){$ \bf u$}
 \put(-250,130){$\bf  P(u)$}
 \caption{$P(u)$ vs $u$ plot corresponding to $\lambda_0$ with $\tilde \kappa = -0.5$ for $N=200$ points}
\label{sp5}
\end{minipage}
\hspace{0.1in}
\begin{minipage}{20pc}
 \includegraphics[width=20pc]{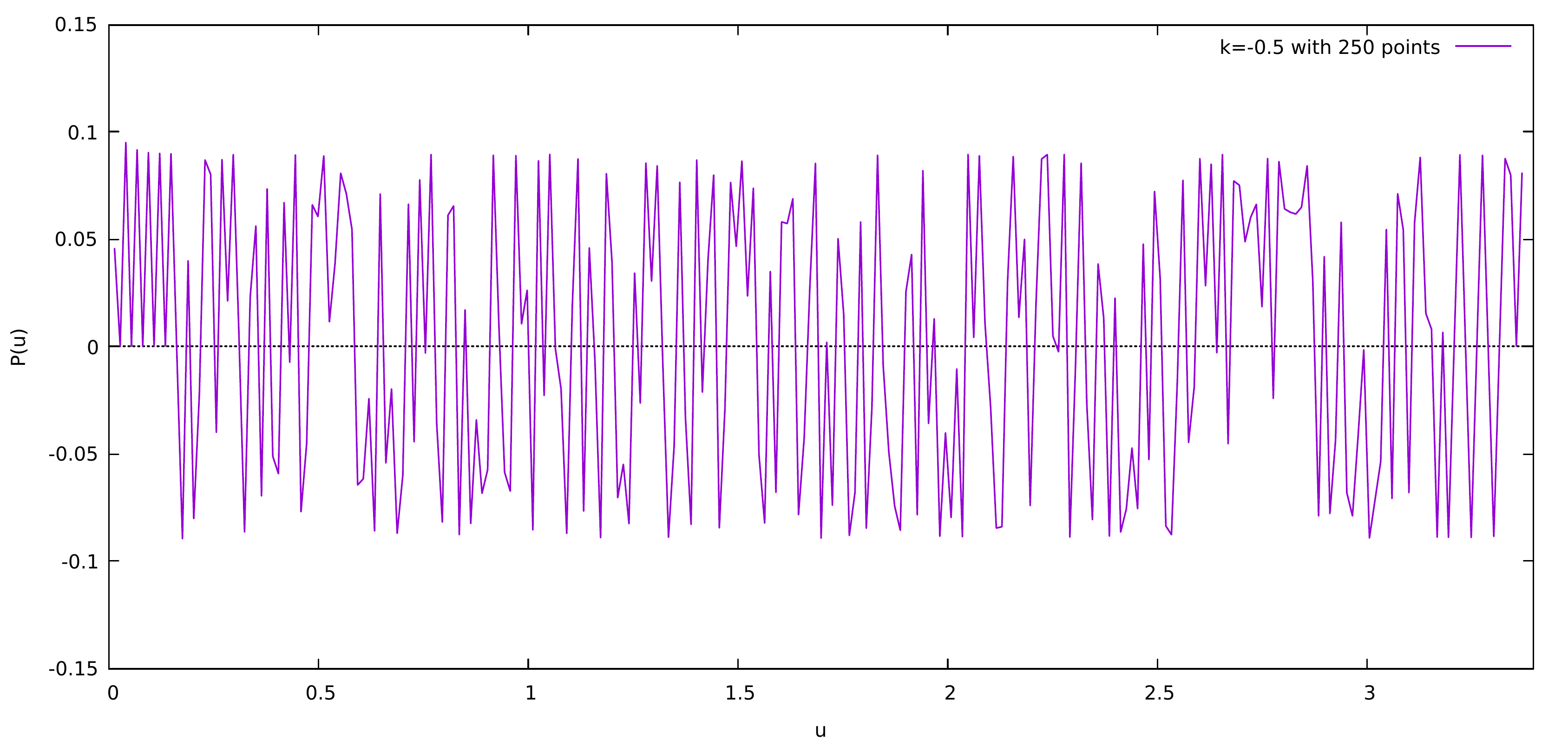}
 \put(-85,-10){$\bf u$}
 \put(-250,130){$\bf P(u)$}
 \caption{$P(u)$ vs $u$ plot corresponding to $\lambda_0$ with $\tilde \kappa = -0.5$ for $N=250$ points}
\label{sp6}
\end{minipage}
\end{figure}
\end{center}
\noindent  We also show some of the solutions $P(u)$ corresponding to $\lambda_0$. In figures (\ref{sp1})-(\ref{sp6}) plots of $P(u)$ vs $u$ for different cases are displayed. Figures (\ref{sp1}) and (\ref{sp2}) represent $P(u)$ vs $u$ plot for $\tilde \kappa = -0.05$ and for $N=200$ and $N=250$ points respectively. Similarly figures (\ref{sp3}) and (\ref{sp4}) are for $\tilde \kappa = -0.1$ and for $N=200$ and $N=250$ points respectively. So are figures  (\ref{sp5}) and (\ref{sp6}) for $\tilde \kappa = -0.5$ and $N=200$ and $N=250$ points respectively. We can see clearly that all of them are oscillatory  finite everywhere, with no divergences) in nature. 
\subsection{The full solution}
\noindent In the previous section we have mainly discussed the solution of the perturbation equation (\ref{pert3}) i.e $P(u)$. Let us now  discuss the nature of $\phi(\tilde \tau, \tilde \sigma)$ i.e the ansatz in eqn (\ref{fs1}). From $\phi(\tilde \tau, \tilde \sigma)$ we construct the normal deformations $\delta x^i = n_{(\alpha)}^{i} \phi^{(\alpha)}(\tilde \tau, \tilde \sigma)$. The normal $n^i$ is shown in the equation \ref{nrm2}. Using eqn \ref{th4} and \ref{scn} we can write the normal (eqn \ref{nrm1}) in the following form 
\begin{equation}
\begin{split}
n^i &= \Big(\frac{\sqrt{\tilde \kappa (1+\tilde \kappa)} ~\cn u}{\sqrt{-1 - \tilde \kappa ~\cn^2 u + \tilde \kappa^2 ~\sn^2 u}},    -\frac{\sqrt{\tilde \kappa}}{\sqrt{-\cn^2 u - \tilde \kappa^2 \sn^2 u}}, \\ & ~~~~~~~~\frac{\cn u}{\cn^2 u - \tilde \kappa ~\sn^2 u}~\frac{\sqrt{1+\tilde \kappa}}{\sqrt{1+\tilde \kappa ~\cn^2 u - \tilde \kappa^2 ~\sn^2 u}}   \Big)
\end{split}
\end{equation}
The normals for three different $\kappa$ ($-0.05, ~-0.1, ~-0.5$) are shown in the following figures (\ref{n1}), (\ref{n2}) and (\ref{n3}). We can see from these plots that the normals are oscillatory in nature. We have already shown in the previous section that $P(u)$ is oscillatory in nature. To get a notion about the full solutions we need to include another quantity -- $e^{i  \beta \tilde \tau}$ which depends on the worldsheet time $\tilde \tau$, as shown in eqn \ref{fs1}. We can see from tables \ref{T1}, \ref{T2} and \ref{T3} that $\lambda_0$ is negative for all the cases and $\lambda_0 = \tilde \beta^2$ and $\tilde \beta^2 = \frac{\beta^2}{1+ \tilde \kappa}$. 
\begin{center}
\begin{figure}[h] 
 \begin{minipage}{10pc}
 \includegraphics[width=10pc]{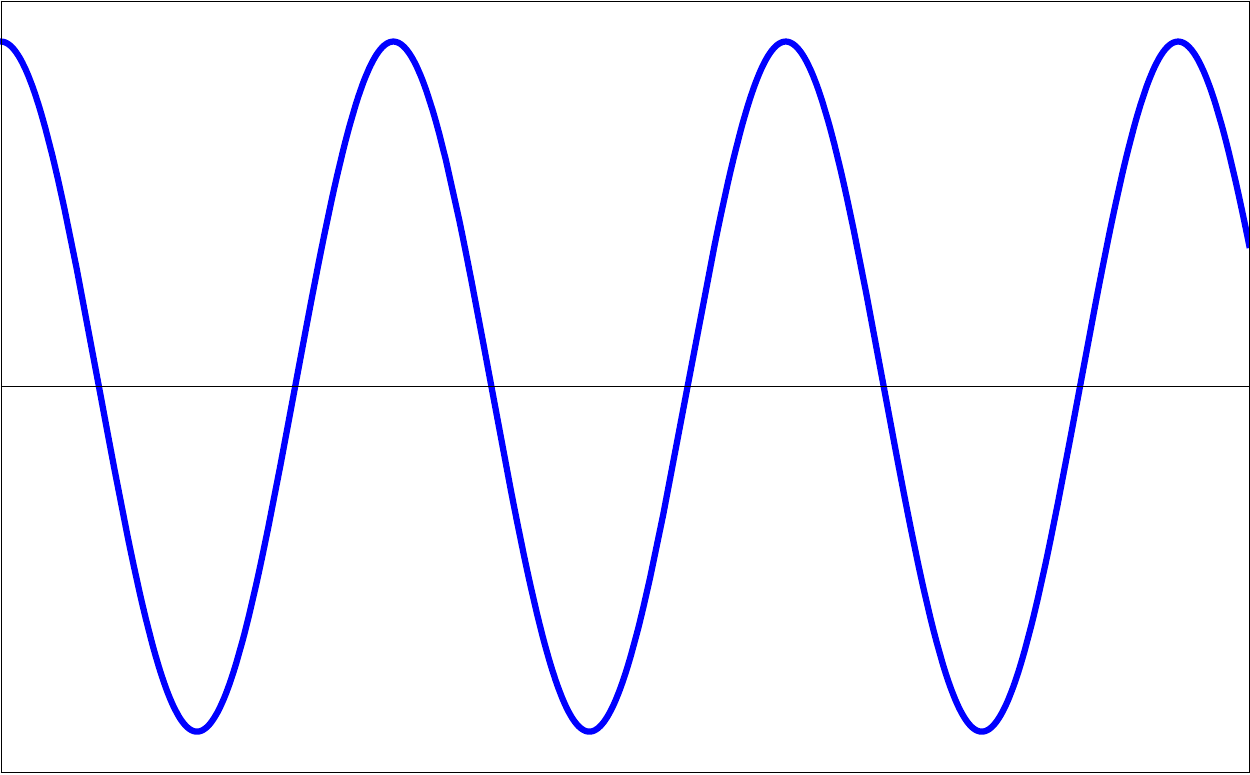}
 \put(-55,-10){$\bf u$}
 \put(-140,50){$\bf  n^1$}
\label{l1}
\end{minipage}
\hspace{0.2in}
\begin{minipage}{10pc}
 \includegraphics[width=10pc]{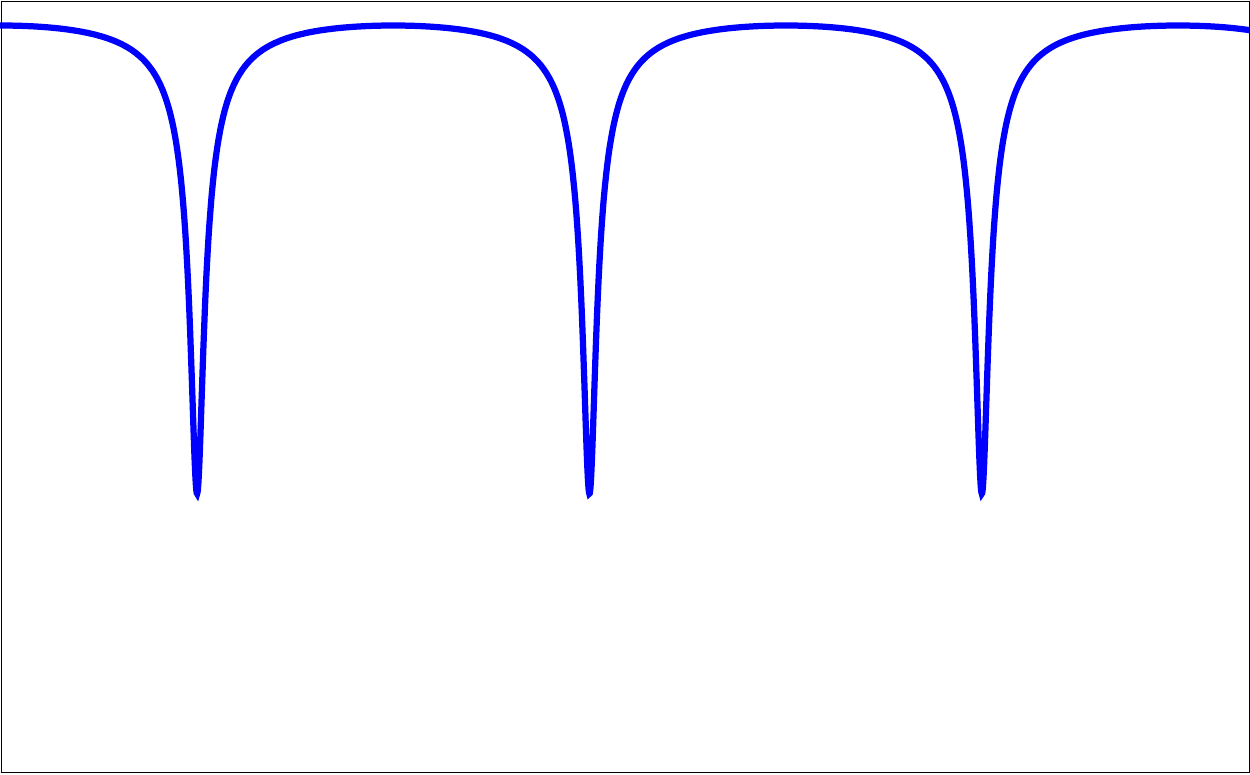}
 \put(-55,-10){$\bf u$}
 \put(-135,50){$\bf n^2$}
 \label{l2}
\end{minipage}
\hspace{0.2in}
\begin{minipage}{10pc}
 \includegraphics[width=10pc]{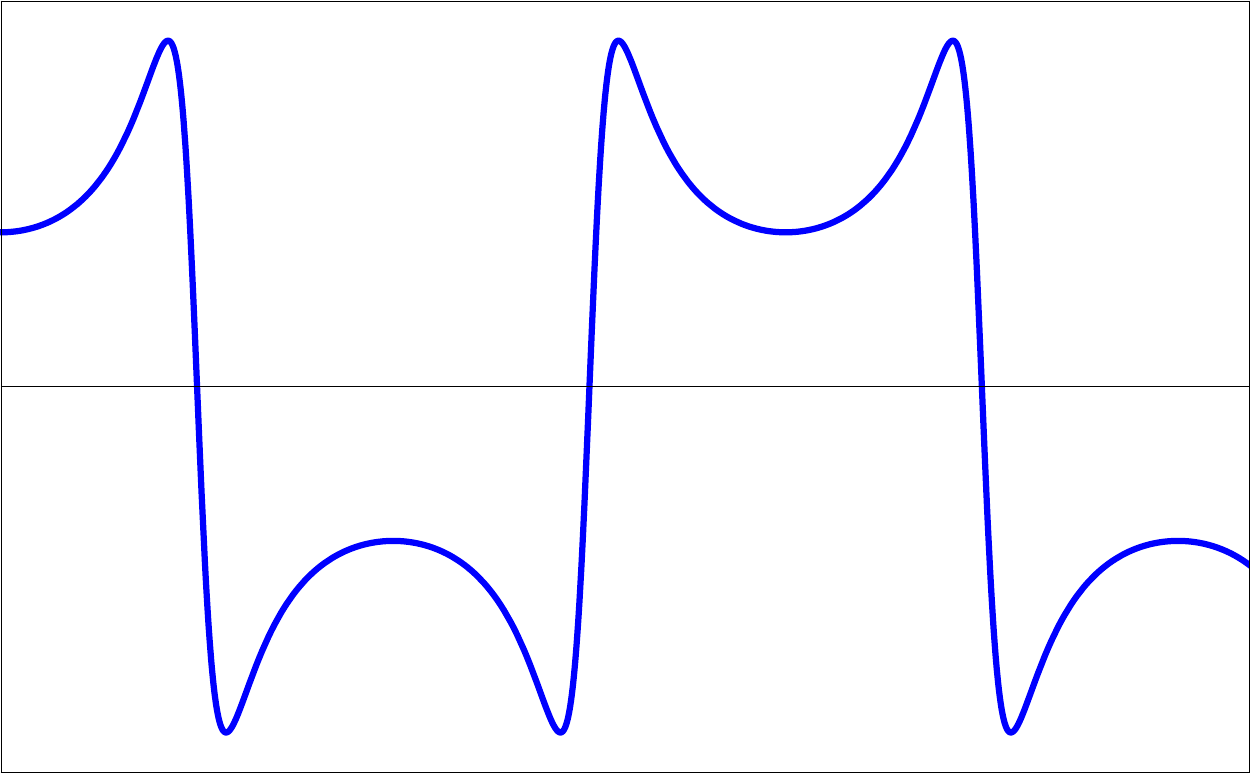}
 \put(-55,-10){$\bf u$}
 \put(-135,50){$\bf n^3$}
\label{l3}
\end{minipage}
\caption{Three components of the normal for $\tilde \kappa = -0.05$ }
\label{n1}
\end{figure}
\end{center}

\begin{center}
\begin{figure}[h] 
 \begin{minipage}{10pc}
 \includegraphics[width=10pc]{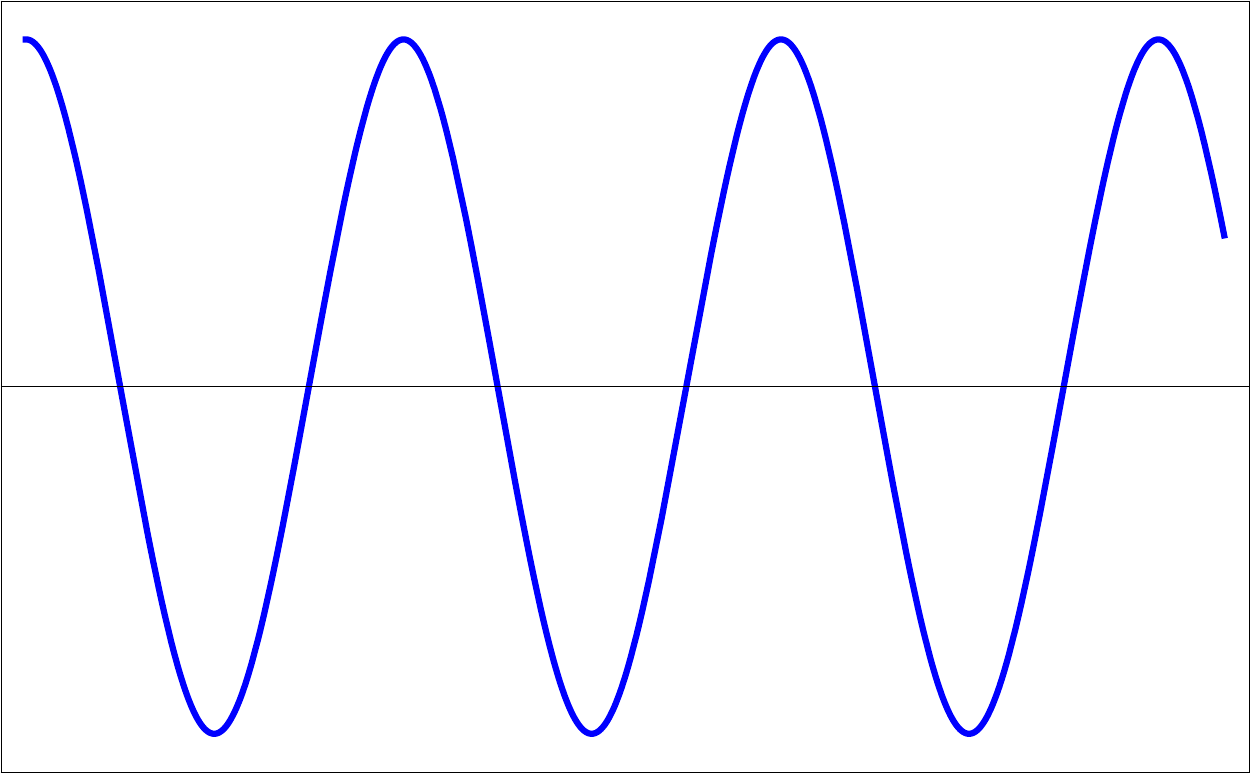}
 \put(-55,-10){$\bf u$}
 \put(-140,50){$\bf  n^1$}
\label{l1}
\end{minipage}
\hspace{0.2in}
\begin{minipage}{10pc}
 \includegraphics[width=10pc]{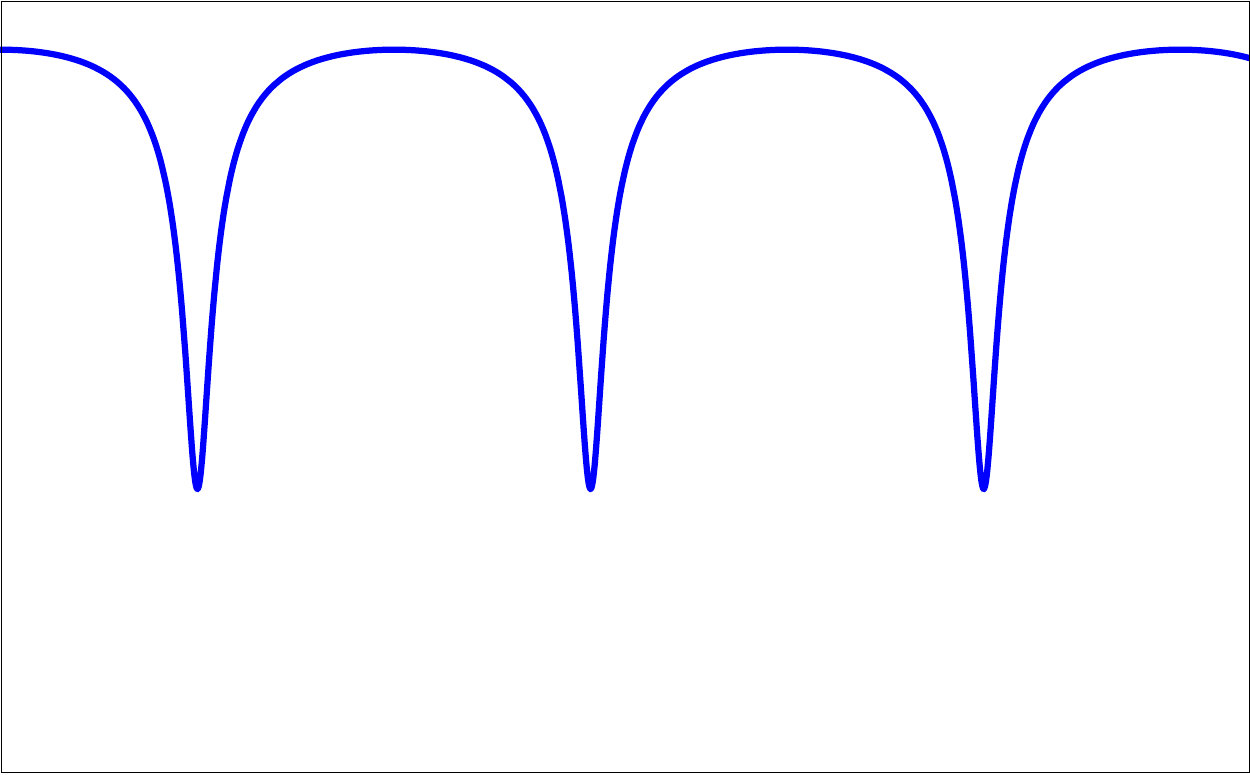}
 \put(-55,-10){$\bf u$}
 \put(-135,50){$\bf n^2$}
 \label{l2}
\end{minipage}
\hspace{0.2in}
\begin{minipage}{10pc}
 \includegraphics[width=10pc]{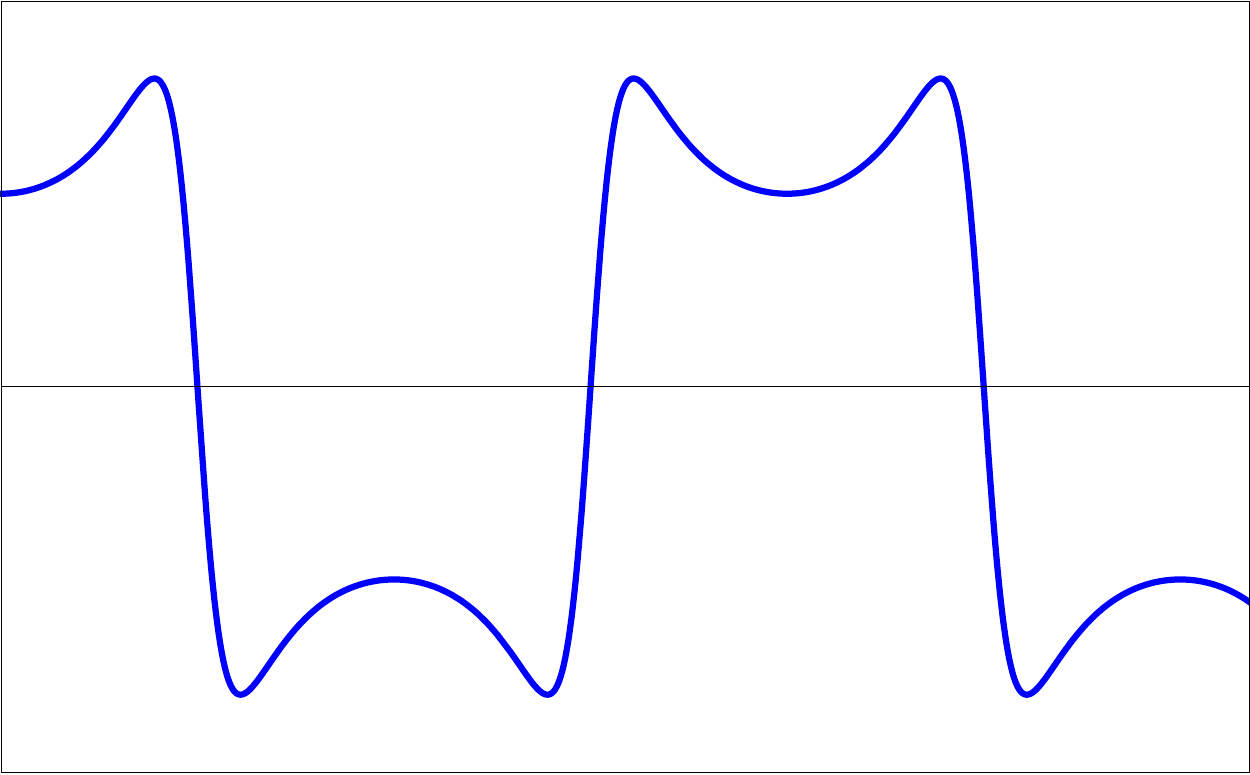}
 \put(-55,-10){$\bf u$}
 \put(-135,50){$\bf n^3$}
\label{l3}
\end{minipage}
\caption{Three components of the normal for $\tilde \kappa = -0.1$ }
\label{n2}
\end{figure}
\end{center}

\begin{center}
\begin{figure}[h] 
 \begin{minipage}{10pc}
 \includegraphics[width=10pc]{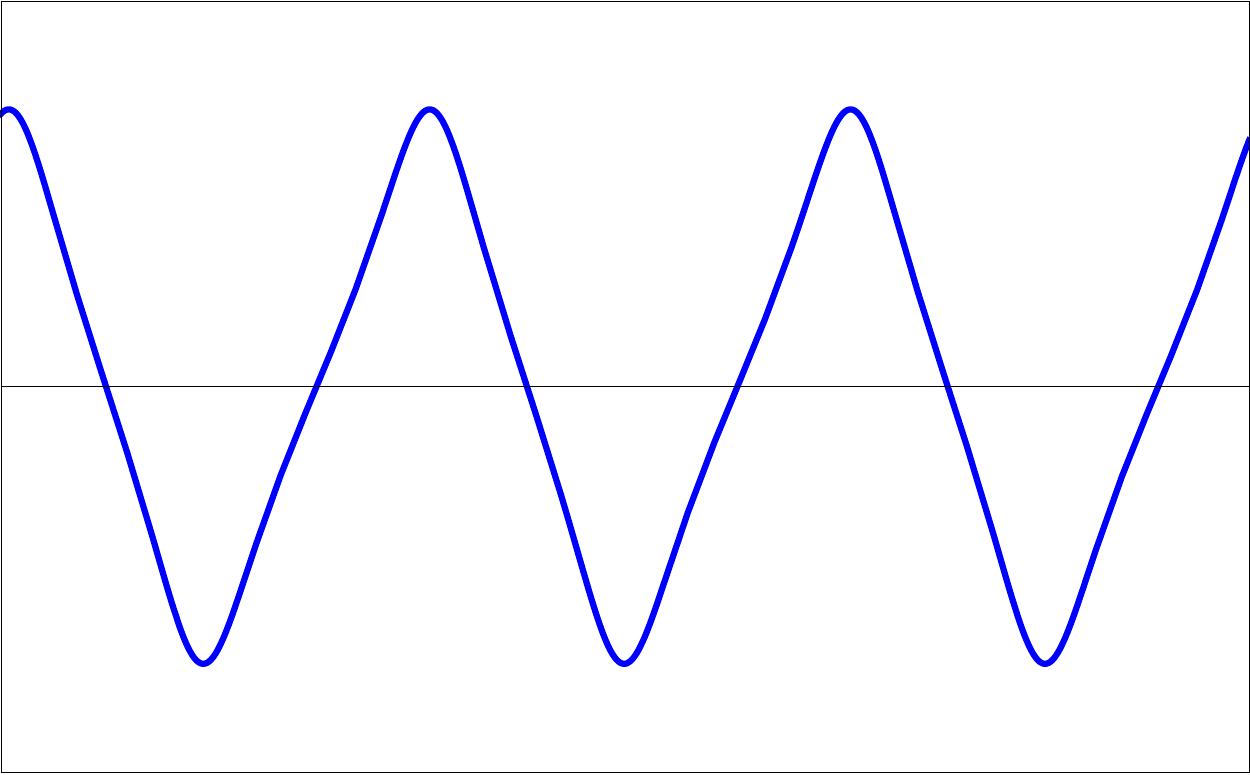}
 \put(-55,-10){$\bf u$}
 \put(-140,50){$\bf  n^1$}
\label{l1}
\end{minipage}
\hspace{0.2in}
\begin{minipage}{10pc}
 \includegraphics[width=10pc]{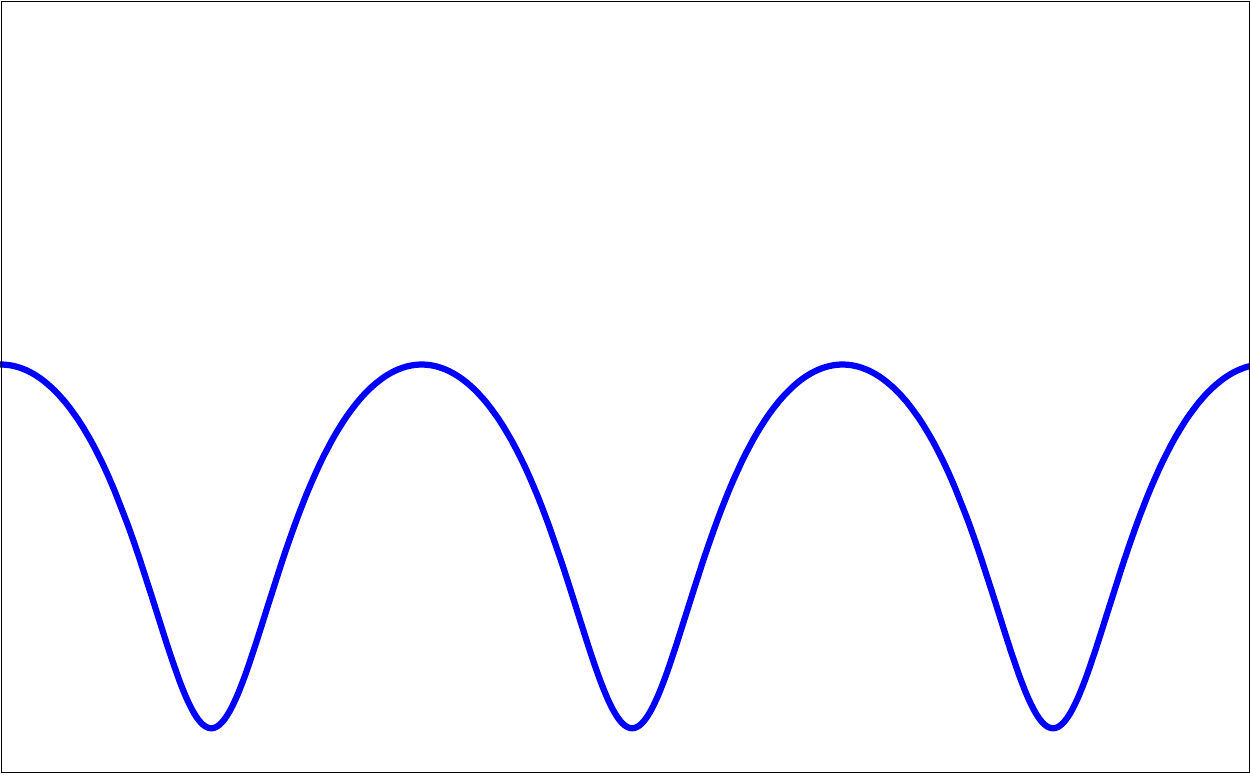}
 \put(-55,-10){$\bf u$}
 \put(-135,50){$\bf n^2$}
 \label{l2}
\end{minipage}
\hspace{0.2in}
\begin{minipage}{10pc}
 \includegraphics[width=10pc]{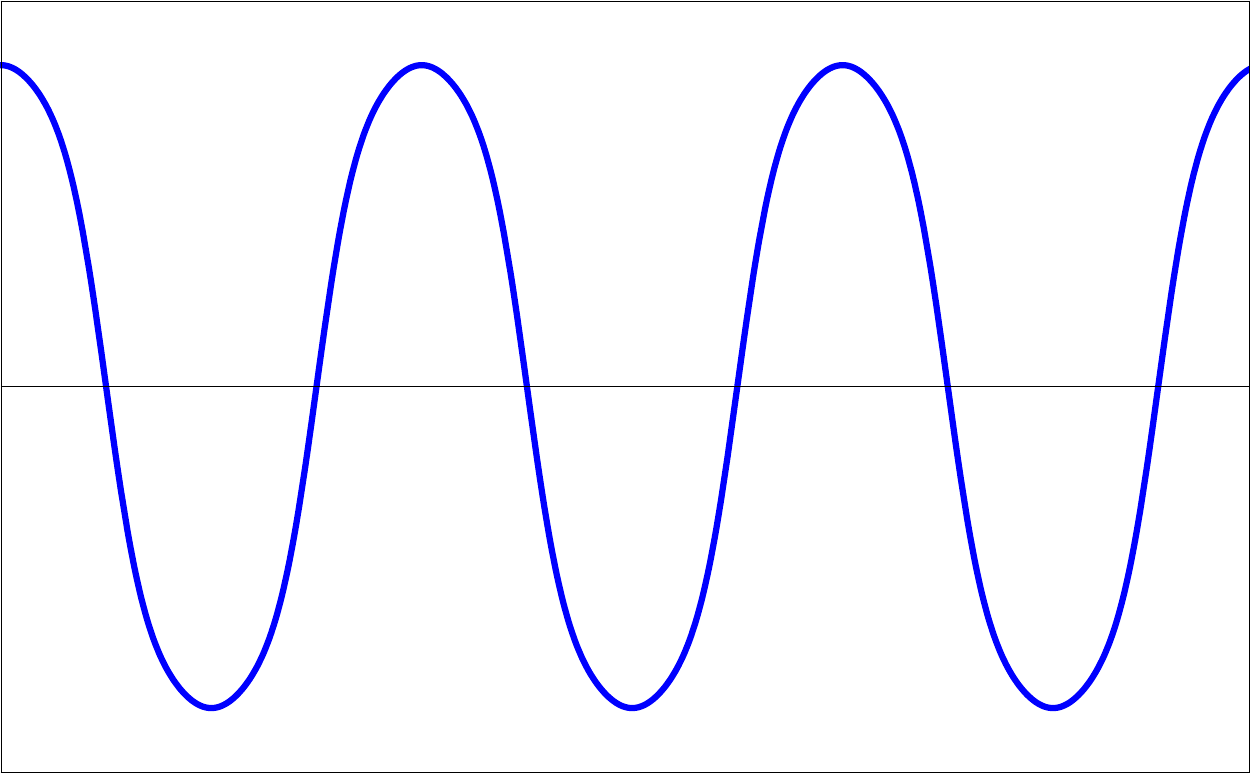}
 \put(-55,-10){$\bf u$}
 \put(-135,50){$\bf n^3$}
\label{l3}
\end{minipage}
\caption{Three components of the normal for $\tilde \kappa = -0.5$ }
\label{n3}
\end{figure}
\end{center}
\noindent Since $1+\tilde \kappa$ is always positive $\beta^2 < 0$. So $\beta = \pm i\eta$ with $\eta>0$. So depending on the sign of $\beta$ we have two cases: 1) When $\beta$ is positive i.e $\beta = +i\eta$ we have the solution like $\delta x^i = \Big(\epsilon~P(u)~n^i \Big)e^{-\eta \tilde \tau}$ which means the perturbation will decreasing exponentially with worldsheet time $\tilde \tau$ and we have decaying mode. 2) When $\beta$ is negative i.e $\beta = -i\eta$ we have the solution like  $\delta x^i = \Big(\epsilon~P(u)~n^i \Big)e^{\eta \tilde \tau}$ which means now we have a exponentially decaying perturbations with respect to worldsheet time. In these two cases  $\Big(\epsilon~P(u)~n^i \Big)$ remains oscillatory as  we have noted above.

\subsection{Analytic solution}
\noindent Here we develop an analytic solution of eqn \ref{pert3} in a certain limit. We mainly focus on the small $\tilde \kappa$ limit  where the potential $V(u)$ reduces to following form,
\begin{equation}
V(u) \equiv \frac{(1-\tilde \kappa)
 (1 + \tilde \kappa ~\sn^2 u)}{(1 + \tilde \kappa) (1 - \tilde \kappa ~\sn^2 u)} \approx \frac{1 - \tilde \kappa}{1 + \tilde \kappa} \Big( 1 + 2 \tilde \kappa \sin^2 u \Big) \equiv \tilde V(u)
\end{equation}
\begin{center}
\begin{figure}[h] 
 \begin{minipage}{15pc}
 \includegraphics[width=15pc]{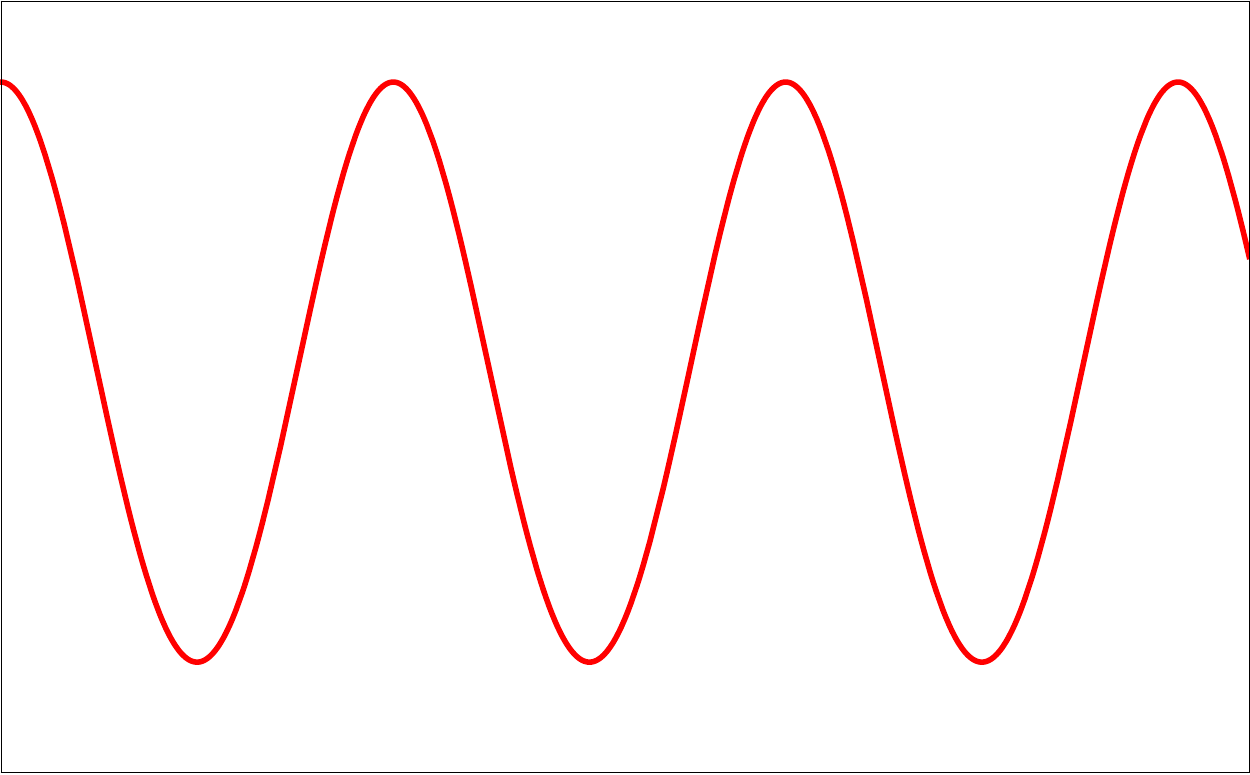}
 \put(-85,-10){$ \bf u$}
 \put(-210,110){$\bf  V(u)$}
 \caption{Plot of original potential $V(u)$ for $\tilde \kappa = -0.05$}
\label{v5}
\end{minipage}
\hspace{0.5in}
\begin{minipage}{15pc}
 \includegraphics[width=15pc]{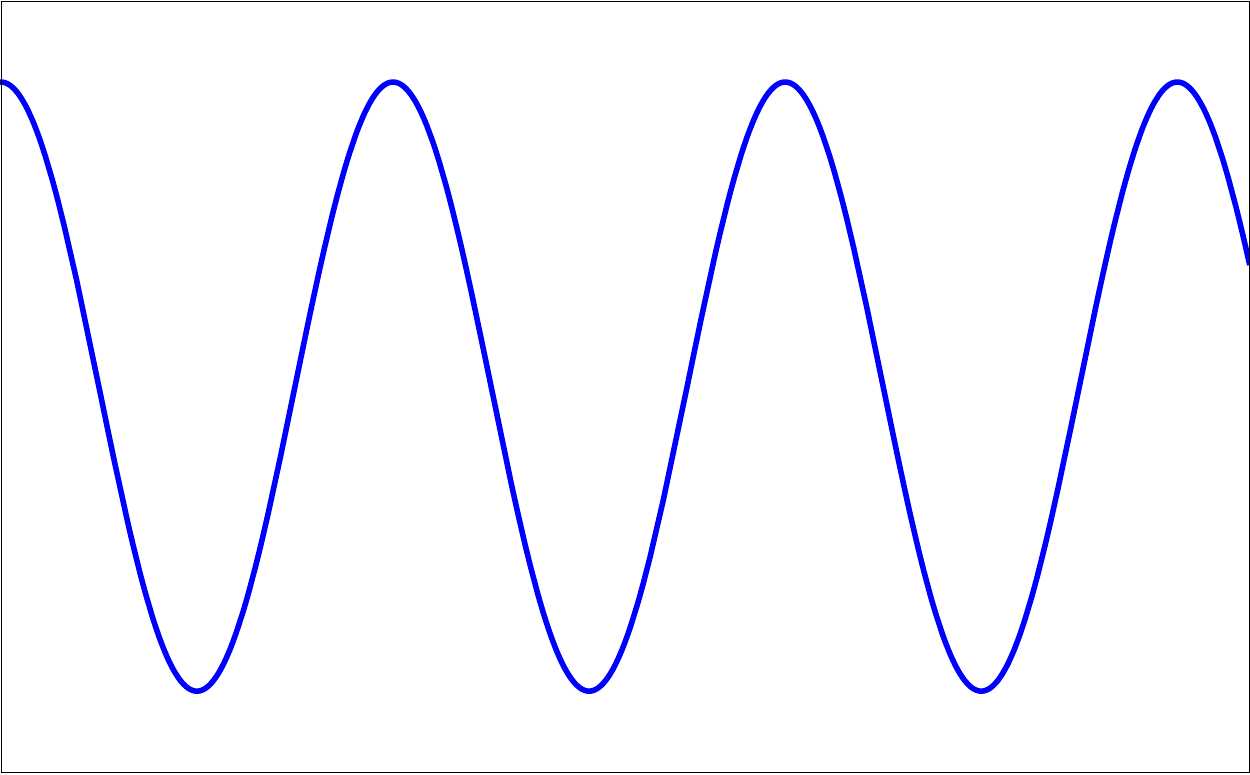}
 \put(-85,-10){$\bf u$}
 \put(-210,110){$\bf \tilde V(u)$}
 \caption{Plot of reduced potential $\tilde V(u)$ for $\tilde \kappa = -0.05$}
\label{v6}
\end{minipage}
\end{figure}
\end{center}
 We consider only the linear order terms in $\tilde \kappa$ and have neglected higher orders. We know at $\tilde \kappa = 0$, $\sn ~u = \sin u$. Thus, at very small value of $\tilde \kappa$ ($\tilde \kappa = -0.05$) $\sn~u$ behaves almost like a $\sin$ function i.e $\sn~(u,\tilde\kappa^2) \equiv \sin u$. We have shown in figure (\ref{v5}) the original potential $V(u)$ and in figure (\ref{v6}) the approximated potential $\tilde V(u)$ for $\tilde \kappa = -0.05$. We can see that both of them are of identical nature which confirms the validity of our approximation.
\noindent  Using the approximated potential $\tilde V(u)$ the perturbation equation \ref{pert3} will reduce to the following form,
\begin{equation}
\frac{d^2 P}{du^2} + \Big(\tilde \beta^2 + \frac{1 - \tilde \kappa}{1 + \tilde \kappa} ( 1 + 2 \tilde \kappa \sin^2 u)  \Big) P = 0
\end{equation}
After some algebraic manipulation one can write the above equation in the following final form
\begin{equation}
\frac{d^2 P}{du^2} + \Big(a - 2q \cos 2u  \Big) P = 0
\label{pert6}
\end{equation}
where the parameters $a$ and $2 q$ are defined as 
\begin{equation}
a = \tilde \beta^2 + \frac{1-\tilde \kappa}{1+\tilde \kappa} + \frac{\tilde \kappa (1-\tilde \kappa)}{1+\tilde \kappa} \, \, \, , \, \,\, 2 q = \frac{\tilde \kappa (1 - \tilde \kappa)}{1+\tilde \kappa} 
\label{pmtr}
\end{equation}

\begin{center}
\begin{figure}[h] 
 \hspace{0.4in}
 \includegraphics[width=25pc]{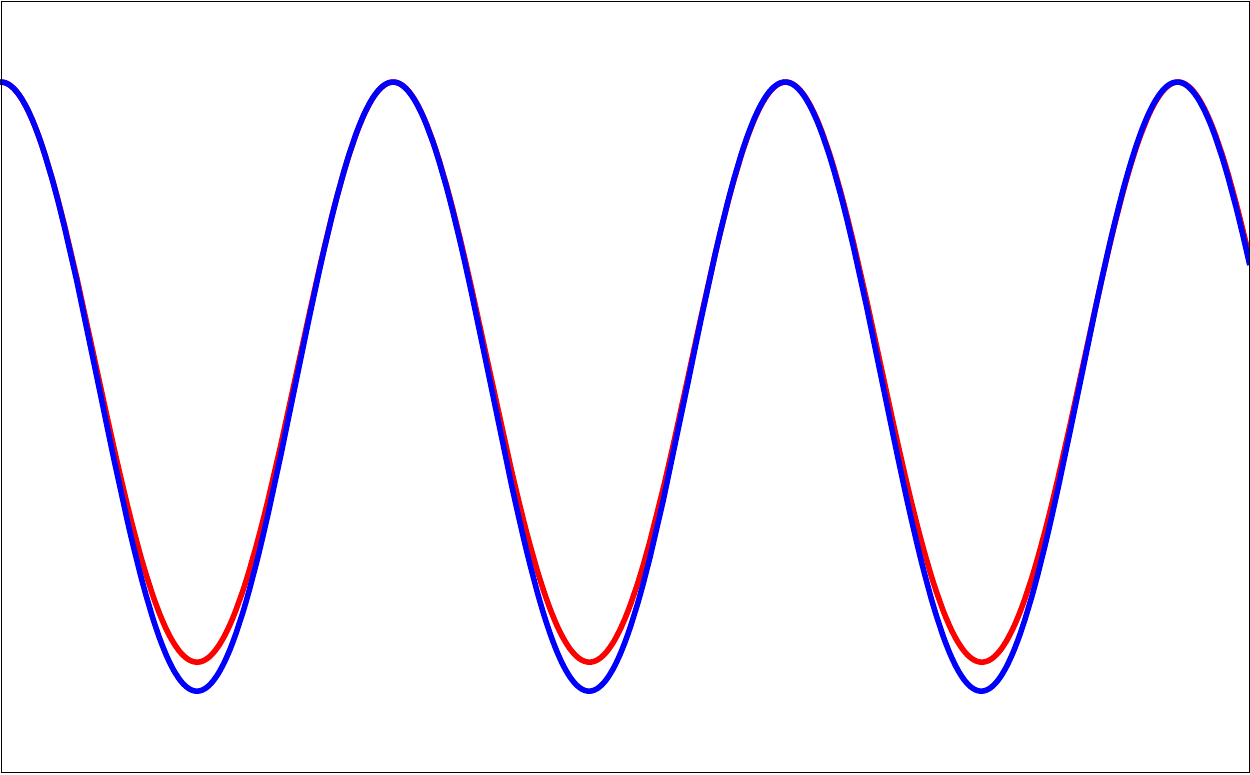}
 \put(-85,-10){$ \bf u$}
 \put(-330,110){$\bf  V(u)$}
 \caption{Plot of original potential $V(u)$ (red) with the reduced one $\tilde V(u)$ (blue) for $\tilde \kappa = -0.05$}
\label{v5}
\end{figure}
\end{center}

Equation \ref{pert6} is the {\bf Mathieu equation}.  We know the solutions of eqn \ref{pert6} analytically as well as numerically.
\noindent A general solution of equation \ref{pert6} will be
\begin{equation}
P(u) = \sum_{m=0}^\infty \Big(A_m \cos m u + B_m \sin m u  \Big)
\end{equation}
with $B_0 = 0$. One can clearly see that the solution is oscillatory in nature. First few roots/characteristic values of this equation can be calculated as power series of parameter $q$
\begin{eqnarray}
a_0(q) = -\frac{q^2}{2} + \frac{7q^4}{128} - \frac{29q^6}{2304} + ... \\
a_2(q) = 4 + \frac{5 q^2}{12} - \frac{763 q^4}{13824} + ..... \\
a_4(q) = 16 + \frac{q^2}{30} + \frac{433 q^4}{864000} + ... 
\end{eqnarray}
First few eigenvalues/roots are shown in table \ref{T10}.  Roots almost match and both the cases follow the nature that $a_r \sim r^2$ as $q \to 0$. So, from table \ref{T10} we can say that our numerical method is appropriate and is tested here via an analytic approximation.
Let us calculate $\tilde \beta^2$ from the analytical result. It turns out to be 
\begin{equation}
\tilde \beta^2 = - \frac{1-\tilde \kappa}{1+\tilde \kappa} - \frac{\tilde \kappa (1-\tilde \kappa)}{1+\tilde \kappa} - a_0 = -1.05038
\end{equation}
Which from the numerics is $\tilde \beta^2 = -1.05039$. From table \ref{T1} one can see that the corresponding value of $\tilde \beta^2 = -1.05231979$. In table \ref{T10} we have shown the comparison between the analytical and numerical values for some more values of $\tilde\beta^2$. These values match quite well with the original un-approximated case of table \ref{T1}. We believe this confirms the validity of  the numerical results discussed in the previous section.
\begin{table}
\caption{Characteristic values calculated analytically and numerically}\label{TT5}
\begin{center}
\begin{tabular}{|c|c|c|} \hline 
$a$  & $ {\rm analytic ~~value}$ & $ {\rm numerical~~value}$  \\ \hline
$a_0$ & $-0.00038172$ & $-0.000387367$ \\
$a_2$ & $4.00032$ & $3.94217$ \\
$a_4$ &  $16.00$   & $15.0859$ \\

\hline
\end{tabular}
\label{T10}
\end{center}
\end{table}

\begin{table}
\caption{Comparison of $\tilde\beta^2$ values}\label{TT6}
\begin{center}
\begin{tabular}{|c|c|c|} \hline 
$\tilde \beta^2$  & $ {\rm analytic ~~value}$ & $ {\rm numerical~~value}$  \\ \hline
$\tilde\beta^2_0$ & $-1.05038$ & $-1.05039$ \\
$\tilde\beta^2_2$ & $2.95032$ & $2.89217$ \\
$\tilde\beta^2_4$ &  $14.95$   & $14.0359$ \\

\hline
\end{tabular}
\label{T10}
\end{center}
\end{table}


\section{Concluding remarks}
\noindent In this paper, first we have shown that the giant magnons in $2+1$ dimensional $\mathbb R \times S^2$ background are stable against normal perturbations. Using the Jevicki-Jin embedding we were able to reduce the perturbation equation into the wave equation in Minkowski background.
 The stability is 
demonstrated through the finite, zero mode deviations from the original profile under small deformations.  We have
shown this explicitly by solving the perturbation equations and by evaluating the effect of perturbations on the embedding.

\noindent Next we have studied normal deformations for single spike solutions in $\mathbb R \times S^2$. We have solved the perturbation equation numerically by using finite difference method and discussed the stability issue extensively through plots. In the end we have shown that under certain approximation the perturbation eqn reduced to the familiar Mathieu eqn. Here we have used our previous numerical method (based on the finite difference method) and have shown that the numerical result matches quite well with the analytical result. This agreement justifies the validity of the numerical results we have discussed so far. We have shown that single spike solution in this background is  largely stable modulo a particular case which we have mentioned in the previous sections.

\noindent The giant magnons in $2+1$ dimensions can easily be extended to $3+1$ dimensions (i.e. for $\mathbb R \times S^3$) and hence it will be interesting to study their stability under similar normal perturbations.

\noindent A further study could be to look at the effect of the perturbations in the 
dual field theory side. It will be interesting to know how these perturbations are eventually related to the 
dual field theory operators.

\section*{Acknowledgements}
\noindent SB would like to thank Dr. Monodeep Chakraborty for several stimulating discussions regarding the numerical method that has been used here extensively.  
\begin{appendices}

\section{Finite difference method}

\noindent Here we briefly introduce the method we have used to solve the perturbation equation (\ref{pert3}) in the context of single spike. The finite-difference methods are a class of numerical techniques for solving differential equations by approximating derivatives with finite differences. Let us consider an interval and divide it into $n$ - sub-intervals each with size $\Delta$.  According to this method one can write the quantity $\frac{d^2 P}{du^2}$ in the following discretized way
\begin{equation}
\frac{d^2 P}{du^2} = \frac{1}{\Delta^2}\Big(P(u_{i+1}) - 2 P(u_{i}) + P(u_{i-1}) \Big)
\label{a1}
\end{equation}
where $P(u_i)$ is the value of the function at $i$-th point. So using eqn \ref{a1} one can write eqn \ref{pert3} in the following set of equations
\begin{align}
\frac{1}{\Delta^2}\Big(P_{2} - 2 P_{1} + P_{0} \Big) + V_1 P_1 &= \lambda P_1 \\
\frac{1}{\Delta^2}\Big(P_{3} - 2 P_{2} + P_{1} \Big) + V_2 P_2 &= \lambda P_2\\
\frac{1}{\Delta^2}\Big(P_{4} - 2 P_{3} + P_{2} \Big) + V_3 P_3 &= \lambda P_3\\
&\vdots   \nonumber        \\ 
\frac{1}{\Delta^2}\Big(P_{n} - 2 P_{n-1} + P_{n-2} \Big) + V_{n-1} P_{n-1} &= \lambda P_{n-1}
\end{align}

\noindent Here $P_i \equiv P(u_i)$, $V_i \equiv V(u_i)$ and $\lambda = -\tilde \beta^2$. Since $P(u)$ is periodic so $P_0 = P_n$. Let us consider a special case of $n=4$ here. We get the following set of equations
\begin{align}
\frac{1}{\Delta^2}\Big(P_{2} - 2 P_{1} + P_{4} \Big) + V_1 P_1 &= \lambda P_1 \\
\frac{1}{\Delta^2}\Big(P_{3} - 2 P_{2} + P_{1} \Big) + V_2 P_2 &= \lambda P_2\\
\frac{1}{\Delta^2}\Big(P_{4} - 2 P_{3} + P_{2} \Big) + V_3 P_3 &= \lambda P_3\\
\frac{1}{\Delta^2}\Big(P_{1} - 2 P_{4} + P_{3} \Big) + V_4 P_4 &= \lambda P_4
\end{align}
We have used the fact $P_0 = P_4$ and $P_5=P_1$ due to periodicity. Above set of equations can be written in the matrix form as follows 
\begin{equation}
\bf M ~P = \lambda ~P
\label{m1}
\end{equation}
where matrix $\bf M$ is the following form
\begin{equation}
{\bf M} =  \begin{pmatrix} -\frac{2}{\Delta^2} +V_1 & \frac{1}{\Delta^2} & 0 & \frac{1}{\Delta^2} \\ \frac{1}{\Delta^2} & -\frac{2}{\Delta^2} +V_2 & \frac{1}{\Delta^2} & 0 \\ 0 & \frac{1}{\Delta^2} & -\frac{2}{\Delta^2} +V_3 & \frac{1}{\Delta^2} \\ \frac{1}{\Delta^2} & 0 & \frac{1}{\Delta^2} & -\frac{2}{\Delta^2} +V_4 
\end{pmatrix}
\end{equation}
where $\bf P$ is $4 \times 1$ column matrix of ${\bf P }= \Big(P_0, ~P_1, ~P_2, ~P_3 \Big)$. 
\noindent We have considered three cases $\tilde \kappa = -0.05, ~-0.1 ~\& -0.5$ and compute the eigenvalues ($\lambda$) and eigenfunctions (${\bf P}$) for certain number of points ($150,~200,~250$) for each cases in section \ref{subsol}.  

\end{appendices}


\end{document}